\documentclass[fleqn,usenatbib]{mnras}

\usepackage{mathtools}
\usepackage{txfonts}

\usepackage[T1]{fontenc}
\usepackage{threeparttable}
\usepackage{xspace}

\DeclareRobustCommand{\VAN}[3]{#2}
\let\VANthebibliography\thebibliography
\def\thebibliography{\DeclareRobustCommand{\VAN}[3]{##3}\VANthebibliography}

\usepackage{graphicx}	

\usepackage[usenames,dvipsnames]{xcolor}
\usepackage{multirow}

\newcommand{\purple}[1]{\textcolor{purple}{#1}}

\newcommand{\blue}[1]{\textcolor{blue}{#1}}

\newcommand{\mpeak}{M_{\rm peak}}
\newcommand{\apeak}{\rm a_{\rm Mpeak}}

\newcommand{\vpeak}{V_{\rm peak}}
\newcommand{\vvirlong}{V_{\rm{vir} \rm{\ (calculated \ from \ M_{\rm peak})}}}
\newcommand{\vvir}{V_{\rm vir}}
\newcommand{\vmax}{V_{\rm max}}
\newcommand{\tinfall}{t_{\rm infall}}

\newcommand{\hMpc}{ h^{-1}{\rm Mpc} }
\newcommand{\ihMpcC}{ h^{3}{\rm Mpc}^{-3}}
\newcommand{\hMsun}{ h^{-1}{\rm M_{ \odot}}}

\newcommand{\mhalo}{M_{\rm halo}}
\newcommand{\msub}{M_{\rm sub}}
\newcommand{\tmpeak}{\rm a_{\rm Mpeak}}

\newcommand{\sig}{\sigma_{8}}
\newcommand{\OmM}{\Omega_\mathrm{M}}
\newcommand{\Omb}{\Omega_{\rm b}}

\newcommand{\ns}{{n_{\rm s}}}

\title[SHAMe-SF: clustering of star-forming galaxies]{SHAMe-SF: Predicting the clustering of star-forming galaxies with an enhanced abundance matching model}

\author[S. Ortega-Martinez]{
	Sara Ortega-Martinez,$^{1,2}$\thanks{E-mail: sara.ortega@dipc.org}
	Sergio Contreras,$^{1}$
	Raúl E. Angulo$^{1,3}$
	\\
	$^{1}$Donostia International Physics Center (DIPC), Donostia-San Sebastian, Spain\\
    $^{2}$University of the Basque Country UPV/EHU, Department of Theoretical Physics, Bilbao, E-48080, Spain\\
	$^{3}$IKERBASQUE, Basque Foundation for Science, 48013, Bilbao, Spain
}

\date{Accepted XXX. Received YYY; in original form ZZZ}

\pubyear{2024}

\begin{document}
	\label{firstpage}
	\pagerange{\pageref{firstpage}--\pageref{lastpage}}
	\maketitle
	
	\begin{abstract}
    With the advent of several galaxy surveys targeting star-forming galaxies, it is important to have models capable of interpreting their spatial distribution in terms of astrophysical and cosmological parameters. To address this need, we introduce SHAMe-SF, an extension of the subhalo abundance matching (SHAM) technique designed specifically for analyzing the redshift-space clustering of star-forming galaxies. Our model directly links a galaxy's star formation rate to the properties of its host dark-matter halo, with further modulations based on effective models of feedback and gas stripping. To quantify the accuracy of our model, we show that it simultaneously reproduces key clustering statistics such as the projected correlation function, monopole, and quadrupole of star-forming galaxy samples at various redshifts and number densities. Notably, these tests were conducted over a wide range of scales $[0.6, 30]\hMpc$, using samples from both the TNG300 magneto-hydrodynamic simulation and from a semi-analytical model. SHAMe-SF can also reproduce the clustering of simulated galaxies that fall within the colour selection criteria employed by DESI for emission line galaxies. Our model exhibits several potential applications, including the generation of covariance matrices, exploration of galaxy formation processes, and even placing constraints on the cosmological parameters of the Universe.
	\end{abstract}
	
	\begin{keywords}
		cosmology: theory -- large-scale structure of universe-- galaxies: formation -- galaxies: statistics -- galaxies: haloes
	\end{keywords}
 
	
	
	\section{Introduction} 

A new generation of galaxy redshift surveys (e.g. DESI, EUCLID, J-PAS, 4MOST) will soon measure the position of hundreds of millions of galaxies. An important aspect in properly analysing these surveys is the creation of realistic mock catalogues that resemble the spatial distribution of galaxies in the Universe. These synthetic catalogues enable us to test analysis pipelines, quantify statistical and systematic uncertainties, and explore the impact of various cosmological ingredients. 

The nature of upcoming surveys will pose several challenges in constructing such mock catalogues. Firstly, the unprecedented amount of data will require extremely precise mock catalogues over large cosmological volumes. Secondly, as these surveys will mostly focus on emission line galaxies (ELGs), the mocks require realistic models for star formation rates and metallicity in galaxies. Thirdly, future observations will provide a complementary view on cosmic large-scale structure by delivering overlapping gravitational lensing and galaxy maps. Consequently, this creates the need for a new generation of mock catalogues with accurate predictions for ELGs and their connection with the underlying dark-matter structures. The SHAMe-SF model presented in this paper, aims to achieve this objective. 

One of the most widely employed approaches to model the spatial distribution of galaxies is the Halo Occupation Distribution (HOD) model \citep{Jing:1998a, Benson:2000, Peacock:2000, Berlind:2003, Zheng:2005, Zheng:2007, Guo:2015a}. Basic HOD models appear sufficient to analyse the clustering of stellar-mass or luminosity-selected galaxies \citep{Zehavi:2005,Coil:2006,Zheng:2007}. However, the modelling of star-forming galaxies is more complex \citep{Geach:2012, C13, C17, Gonzalez-Perez:2020, Avila:2020, Alam:2020} since, for instance, those galaxies typically do not follow the same phase-space distribution as dark matter and their abundance does not scale monotonically with halo mass \citep{Orsi:2018,Avila:2020}. Additionally, HODs make several simplifications, such as the lack of assembly bias \citep[e.g.,][]{Sheth:2004, Gao:2005} which can lead to incorrect cosmological inferences \citep{Cuesta-Lazaro:2023,ChavesMontero:2023}. Recently, some of the limitations have been addressed by extending HODs with environmental dependencies and/or secondary halo properties in addition to mass \citep{Hearin:2016, Zehavi:2018, Xu:2021, Hadzhiyska:2022b, Hadzhiyska:2022}. However, the greater number of parameters decreases the predictability of the model, potentially compromising its ability to accurately predict statistical quantities beyond those explicitly tested.

It is possible to obtain a more realistic galaxy distribution by directly assuming a connection between galaxies and subhalos. Specifically, in Subhalo Abundance Matching (SHAM) models, the most massive (luminous) galaxies are assumed to be hosted by the largest subhalos \citep{Vale:2006, Shankar:2006}. This model is notably successful at reproducing the spatial distribution of galaxies selected based on stellar mass in both observational datasets and hydrodynamical simulations \citep[e.g.][]{Conroy:2006, ChavesMontero:2016}. In the case of SFR-selected galaxies, SHAM is relatively successful at high-z ($z \sim 2$), where the fraction of quenched galaxies is small \citep{Simha:2012}. However, its accuracy decreases at low redshift, where subhalo mass is a poor predictor for star formation \citep[see e.g][]{C15}. A potential approach to address this issue is to begin with a standard SHAM and then incorporate an additional subhalo property to model the relationship between SFR and stellar mass \citep{Tinker:2018, Favole:2022}. 

Hydro-dynamical simulations and Semi-Analytical Models (SAM) provide a more sophisticated way to construct mock surveys. These approaches attempt to directly model relevant physical processes (such as star formation, feedback from AGNs and supernova, and metal enrichment) whose free parameters are calibrated by reproducing a set of predefined observations \cite[see][for reviews]{Baugh:2006,Vogelsberger:2020rev}. However, it is still not clear how well hydrodynamical simulations and SAMs reproduce the observed SFR, especially at higher redshifts where observations are scarce (e.g., \citealt{McCarthy:2017,SIMBA} for simulations and \citealt{Wang:2018} for SAMs). Additionally, the large computational cost of these models usually means that only a handful of variations of the underlying physical recipes and free parameters are available. Moreover, the volumes are still usually comparatively smaller than those available to simpler models. 

Ultimately, the physics that regulates star formation and emission lines in galaxies is still uncertain and, almost certainly, its implementation in numerical simulations is too simplistic. This motivates the development of ``Empirical Assembly Models'' which describe the relation between galaxies and their host dark matter subhalos using parametric equations that aim at being an "effective" description of the relevant physics \citep{Moster:2018,Behroozi:2019}. For instance, \texttt{Emerge} \citep{Moster:2018} assumes that the SFR of a galaxy is given by the dark-matter accretion rate times a baryon conversion efficiency but without making a reference to the specific physics responsible. The philosophy of these models is that the effective descriptions can be constrained through observations, which would then help to identify the actual physical processes governing galaxy formation. 

If one aims to only reproduce galaxy clustering, it is possible to further reduce the number of assumptions. An example of this is the SHAMe model presented in \cite{Contreras:2021shame}. The SFR prescription in SHAMe combines the recipes implemented in \texttt{Emerge} with the rank ordering philosophy of the SHAM models. This approach has several advantages. For instance, we can achieve realistic models with only a few free parameters while being fast and computationally cheap. This allows scanning millions of different subhalo-galaxy connections and vary cosmology. 

The SHAMe model is a significant improvement compared to other prescriptions. Importantly, it delivers accurate predictions for the joint distribution of clustering and lensing statistics \citep{C2023:lensing,C23:lensingobs}. However, the predictions for SFR-selected galaxies were much less accurate than for stellar-mass selections. In this work, we present SHAMe-SF, a new galaxy model that improves the SFR prescription of \cite{Contreras:2021shame} to better model the spatial distribution and velocities of the star-forming galaxies to be detected by upcoming large-scale structure surveys. 

To achieve our goal, we first apply machine learning techniques onto a hydrodynamical simulation to identify the most important subhalo properties for predicting the SFR of its hosted galaxy. Using this information, we then build a model that can be applied to subhalos in a gravity-only simulation. The new model populates a simulation with $205\hMpc$ side in just a few seconds. Additionally, it is flexible enough to reproduce the clustering of samples from both hydrodynamic simulation and semi-analytical models. The flexibility and computational efficiency of SHAMe-SF could help to build realistic mocks for star-forming galaxies that span our uncertainty regarding SFR physics and for many different cosmological models. Ultimately, these features will allow SHAMe-SF to be used directly in the interpretation of upcoming galaxy redshift surveys and thus placing constraints on astrophysics and cosmology with such observational data. 

This paper is organised as follows. In \S\ref{sec:simsandrest} we present the simulations and models used in this work. In \S\ref{sec:RF} we employ a Random Forest algorithm to identify the best combination of subhalo properties to predict the spatial distribution of galaxies selected by SFR at $z=0$ and $1$. In \S\ref{sec:Model} we employ those properties to build a model that could be applied to gravity-only simulations. In \S\ref{sec:clusteringall} we validate our model by presenting the best fits to redshift-space clustering, whereas in \S\ref{sec:otherstats} we focus on additional tests. We summarise our findings in \S\ref{sec:conclusions}.  

\begin{table*}
 \begin{threeparttable}[t]
		\begin{tabular}{lll}
			\hline
			\textbf{1. Mass proxies}        & $\msub$      & Current mass of the subhalo             \\
			& $\mpeak$      & Maximum value of $\msub$ through the evolution of the subhalo\\
			& $\mhalo$ &      Mass of the host halo ($\sim \msub$ for centrals)                                  \\
			& $\vmax$       &       Maximum value of $v(r)$ for each subhalo                                  \\
			& $\vpeak$      &     Maximum value of $\vmax$ through the evolution of the subhalo                                    \\ 
            
			\hline
			\textbf{2. Time indicators}    & $\tinfall$    & Last time a subhalo was identified as a central                \\
			& $\tmpeak$     &  Time when the subhalo reached its peak mass ($\mpeak$)                                       \\
			& $t_{\rm Vpeak}$    &  Time when the subhalo reached its peak $\vmax$ ($\vpeak$)                                         \\
			& Age           &     Half-mass formation time of a subhalo                                    \\  
			& Halo age           &  Half-mass formation time of the host halo                                          \\ 
			\hline
			\textbf{3. Accretion}    & $\dot{M}$    &     Difference in $\msub$ between the current and the previous snapshot                 \\
			& $\dot{M}_{\rm{at peak}}$    &   Difference in  $\msub$ between $t_{\rm Mpeak}$ and its previous snapshot                                             \\
			& $\dot{V}$     &   Difference in $\vmax$ between the current and the previous snapshot                                       \\
			& $\dot{V}_{\rm{at peak}}$     &    Difference in  $\vmax$ between $t_{\rm Vpeak}$ and its previous snapshot                                       \\ 
			\hline
			\textbf{4. Environmental} & Linear bias           &     Object-by-object bias \citep{Paranjape:2018, Paranjape:2020}                \\
			&$\alpha_{\rm peak}$            &     Local tidal anisotropy \citep{Ramakrishnan:2019,Zjupa:2020}                                   \\
			& Central distance to halo          &    Distance to the central subhalo (0 for centrals)                                     \\
			& Min central dist to halo          &     Minimum value of the Central distance to halo through the evolution of the subhalo                                   \\
			\hline
			\textbf{5. Other properties} & Angular momentum            &   Modulus of the halo angular momentum                \\
			\textbf{\& ratios} & $\vmax/V_{200,\rm t_{infall}}$           &     Concentration proxy. Ratio between $\vmax$ and $\vvir$ when each subhalo was last a central  \\ 
			& $\vpeak/\vvirlong$           &    Concentration proxy, calculated without the need of the full merger tree.  \\ 
			&$\langle \nu \rangle_{t}$            &       Integrated growth history (\blue{Chaves-Montero et al. in prep.})                                 \\
			&$\msub/\mhalo$           & Ratio between the mass of the subhalo and its host halo          \\
			&$\msub/\mpeak$              &  Ratio between the current mass and the peak mass of a subhalo                                       \\
			&$\vmax/\vpeak$             &   Ratio between the peak mass and the current mass of a subhalo                                      \\
			\hline
			                        
		\end{tabular}

    \end{threeparttable}%
		\caption{Definitions of the subhalo properties considered for the RF.}
		\label{tab:variables}
	\end{table*}

\section{Simulations and galaxy population models}
\label{sec:simsandrest}
In this section, we provide an overview of the hydrodynamical and gravity-only simulations we employ in this work, along with a description of the samples we use to develop and validate SHAMe-SF. Additionally, we introduce the Semi-Analytical Model (SAM) employed to assess SHAMe-SF's capability to replicate the clustering of samples based on an alternate star formation rate (SFR) prescription.

\subsection{TNG300}
\label{sec:TNG}

\subsubsection{Simulations}
    To analyse the connection between subhalo properties and SFR as well as test the performance of SHAMe-SF, we use the magneto-hydrodynamic simulation Illustris-TNG300 \citep{Nelson:2017TNGcolors, TNGb, TNGc, TNGd, TNGe}. 
    
    The TNG300 simulation is part of the "The Next Generation" Illustris Simulations suite, and it is one of the largest publicly available high-resolution hydrodynamical simulations. We use TNG300-1 (TNG300 thereafter), the run with the highest resolution for the largest box, which jointly evolves 2500$^3$ dark matter particles and gas cells on a periodic box of $205\hMpc$ ($\sim$ 300 Mpc) side. The mass of these elements are $3.98\times10^7\,\hMsun$ and $7.44\times10^6\,\hMsun$, respectively. The simulation was run using \texttt{AREPO} \citep{AREPO} with cosmological parameters from \cite{Planck2015}\footnote{$\OmM$ = 0.3089, $\Omb$ = 0.0486, $\sig$ = 0.8159, $\ns$ = 0.9667 and $h$ = 0.6774.}. In addition, we will use the TNG300-1-Dark simulation, the gravity-only counterpart of the TNG300, which contains the same number of dark matter particles (with $4.72\times10^7\,\hMsun$ mass), volume and initial conditions as TNG300. 
    
To match subhaloes in the TNG300-1-Dark to those in the TNG300, we use one of the catalogues provided by the TNG collaboration \citep{Nelson:2015, Springel:2005}, which was created using a bidirectional matching (subhalos are only linked only if they share most of their particles among gravity-only and baryonic runs). At $z = 1$, the completeness is almost 100\% for central subhalos, and $\sim60\%$ for satellites when considering subhalos with $\mpeak > 3\times 10^{10}\,\hMsun$, the mass of a subhalo resolved with 10 particles in the lowest resolution run (similarly at $z=0$).
    
    As shown in \cite{Springel:2018}, TNG300 agrees well with observational estimates of the clustering of stellar-mass selected galaxies in SDSS. Additionally, TNG300 qualitatively reproduces the main sequence and scatter of the stellar mass-SFR relation \citep{Donnari:2019}. TNG300 also reproduces the general trends in the fraction of quenched galaxies. However, it cannot replicate a strong bimodality on SFRs \citep{Donnari:2019, Zhao:2020}. Regarding colours, TNG300 can reproduce the shape of the red sequence and the blue cloud \citep{TNGa} when compared to SDSS data, as well as the bimodality in colour.

	Therefore, TNG300 provides us with a reasonably realistic population of star-forming galaxies for our work. However, we emphasise that our goal is not to perfectly reproduce this simulation, instead, we aim to use it to extract general physical relations between SFRs and subhalo properties. We will use these relations to develop a general empirical model that could be used to describe galaxies in TNG300 but also in other physical galaxy formation models, and ultimately on observations of our Universe.   

\subsubsection{Number density-selected samples}

\begin{table}
\begin{tabular}{lccc}
Number density & SFR [M$_\odot$/Yr] & SFR [M$_\odot$/Yr] & $N_{\rm subhalos}$ \\
$[\ihMpcC]$ &  z = 0 & z =1 & \\
 \hline
$10^{-2}$ & 0.47 & 2.35 & $8.61\times 10^5$ \\
$10^{-2.5}$ & 1.50  & 6.86 & $2.72\times 10^5$ \\
$10^{-3}$ & 3.03  & 13.57 & $8.61\times 10^4$ \\
$10^{-3.5}$ & 5.14 & 22.21  & $2.72\times 10^4$
\end{tabular}
\caption{SFR-selected samples constructed from the TNG300 hydrodynamical simulation at $z = 0$ and $z = 1$ for different number densities. The second and third columns provide the minimum SFR included in the sample, whereas the fourth column shows the total number of subhalos in each sample.}
\label{tab:SFR}
\end{table}
    
We will consider 4 different samples of galaxies at $z=0$ and $1$. These samples correspond to number-density selections in SFR: $\bar{n} = \{10^{-2},10^{-3}, 10^{-2.5}, 10^{-3}\}\,[\ihMpcC]$. The number of objects, as well as the minimum SFR value (in $\rm M_{*}/Yr$), are provided in Table~\ref{tab:SFR}. We note we define our samples in terms of number densities since SHAMe-SF predicts the rank order of star-forming galaxies and not the actual value of their SFR. 

The first three samples match the number densities of those in \cite{Contreras:2021shame}, thus it will allow us to compare directly the 
performance of our model with its predecessor. The 4th sample roughly matches the abundance expected in DESI \citep[e.g.,][]{DESI:2016, Yuan:2022b}, thus we will use it to test the model performance for such survey.     
    
\subsubsection{Colour-selected samples}
\label{sec:ELG}

    As mentioned above, one of the targets of upcoming surveys will be Emission Line Galaxies (ELG). ELGs are detected by measuring the flux of some specific lines, such as the [OII] doublet or H$\alpha$, associated with newly formed stars and their interaction with the surrounding gaseous media.
    
	Since the TNG300 does not provide predictions for specific emissin lines, we will test the performance of our model on a sample with similar properties to DESI's selection of ELGs. Following \cite{Hadzhiyska:2021}, we adopt various magnitude and colour selection criteria \citep[][see Table~\ref{tab:ELG}]{DESI:2016,Raichoor:2020}. 
    We use synthetic SDSS magnitudes accounting for dust obscuration calculated for TNG300 galaxies by \cite{Nelson:2017TNGcolors}. This catalogue is available on the TNG database, and the dust model is outlined in the aforementioned work.
    
    Imposing these selection criteria to TNG300 galaxies, we obtain a number density of $n \approx 0.002\,\ihMpcC$ at $z = 1$ (note that it is between the two intermediate samples described in the previous subsection).
    
\subsection{TNG300-mimic}
    \label{sec:TNGmimic}
    In addition to TNG300-1-Dark, we use a gravity-only simulation with lower resolution, referred to as TNG300-mimic. This will test whether SHAMe-SF can be used with a resolution achievable in large-volume simulations. 
    
    The TNG300-mimic simulation has the same volume and initial conditions as TNG300-1-Dark, but employs only 625$^3$ particles (equivalent to TNG300-3-Dark). It was carried out with an updated version of \texttt{L-Gadget3} \citep{Angulo:2012,Springel:2005}, employed for the Millennium XXL and BACCO simulations \citep{Angulo:2021}. It includes an on-the-fly identification of halos and subhalos through a Friends-of-Friends algorithm \citep[\texttt{FOF}][]{Davis:1985}, and a modified version of SUBFIND \citep{Springel:2001} able to follow the history of subhalos and compute properties which are non-local in time (peak masses and velocities, accretion rates, etc.). Numerical accuracy parameters match those employed in the BACCO simulations \citep[see][]{Angulo:2021}.

Given the identical initial conditions of TNG300-mimic and TNG300 simulations, the comparison between SHAMe and TNG samples is less susceptible to the effects of cosmic variance.

    \subsection{Semi-analytical model}
	\label{sec:sam}
Using TNG300 as a reference assumes a specific prescription for modelling baryonic physics. However, our goal is to develop a model capable of capturing various plausible galaxy formation scenarios rather than being confined to a single simulation. Because of this, we employ a semi-analytical model (SAM) to further validate SHAMe-SF. In this work, we will employ the public version of \texttt{L-Galaxies} \footnote{\url{http://galformod.mpa-garching.mpg.de/public/LGalaxies/}}, developed by the "Munich Group" \citep{White:1991, Kauffmann:1993, Kauffmann:1999, Springel:2001,DeLucia:2004, Croton:2006, DeLucia:2007, Guo:2011, Guo:2013, Henriques:2013,Henriques:2020}, as implemented by \cite{Henriques:2015}. The model was executed with its default parameter set on TNG300-mimic, as in the fiducial SAM model in \cite{C2023:lensing}. Subsequently, we generate galaxy catalogues analogous to those constructed from TNG300 (c.f. Table 2).

    \begin{table}
    \begin{tabular}{ll}
    \hline
    \multirow{3}{*}{Magnitude limits} & $20.0<g<23.6$ \\
     & $r < 23.4$ \\
     & $z < 22.5$ \\ \hline
    \multirow{3}{*}{Colour selection} & $0.3<(r-z)<1.6$ \\
     & $(g-r)<1.15(r-z)-0.15$ \\
     & $(g-r)<-1.2(r-z)+1.6$ \\ \hline
    \end{tabular}
    \caption{Selection criteria used to construct a sample of DESI-like ELGs in the TNG300 hydrodynamical simulation. All magnitudes are computed assuming SDSS filters in the AB system.}
    \label{tab:ELG}
    \end{table}

\section{The relation between star formation rates and dark matter structures}
	\label{sec:RF}

Before attempting to build a physically-motivated model for star-forming (SF) galaxies, we first seek to understand the relation between the properties of simulated galaxies and those of their host DM subhaloes. This correlation might not be trivial due to the diverse array of physical processes at play, including gas cooling and accretion, feedback mechanisms, RAM pressure stripping, and galaxy interactions. In fact, in the literature the SFR has been linked to various DM properties such as host halo mass, halo mass accretion rate, concentration, age, and environment \citep{Wang:2013,Behroozi:2013EQUATION,ChangHoon:2019,Blank:2020,Zjupa:2020}.  
 
Here, we will employ the TNG300 catalogues together with a machine learning algorithm, specifically a Random Forest, to discern the subset of subhalo properties that best predict the spatial distribution of SF galaxies. The following subsections provide details of our approach and demonstrate that the clustering of SFR-selected samples from TNG300 can be accurately predicted using only 3 variables -- namely the subhalo mass and concentration, and the host halo mass.

\subsection{Random Forest}

   Random Forest (RF) algorithms are designed to predict a set of values given a set of input variables. During the training phase, the algorithm builds several decision trees using a random subsample of the input data \citep{Breiman:trees}. At each level of the tree, the algorithm performs a division using the input variable that minimises a given loss function (e.g. squared error). The number of divisions is typically limited to maintain a given number of instances per leaf. Since each tree is built with a random subset of the training data, RF is not prone to overfitting. The final prediction is obtained by walking the tree and averaging the values of data within the final leaf across all random trees \citep{Breiman:2001}. 


Tree-based algorithms have been used to predict (s)SFR based on subhalo properties \citep{Kamdar:2016,Agarwal:2018,deSanti:2022}, yielding reasonably accurate predictions (albeit less precise than for stellar mass). However, since RF trees are built by making divisions using one of the input variables at each step, they are well-suited for exploring the influence of input properties on predictions. Indeed, RFs have been used, for instance, to disentangle the role of secondary variables in the galaxy-halo connection. Here, we will employ RFs to identify the DM variables that best predict the SFR of TNG300 galaxies \citep[see][for a similar approach]{Moster:2021}.

\subsubsection{Training}

Our basic idea is to build a suite of RFs that consider different subsets of DM properties. In particular, we consider those properties listed in Table~\ref{tab:variables}, containing properties computed from halo and subhalo catalogues (e.g. $\mhalo$, $\vmax$), derived from subhalo merger trees (e.g., $\vpeak$, halo age), as well as measured from the large-scale environment (e.g., linear bias, $\alpha_{\rm peak}$). 

In principle, we would like to consider all possible combinations of $N$ subhalo properties. However, this quickly becomes computationally infeasible. As a pragmatic approach, we will only focus on the best-performing set of $N$ variables as a basis for building RFs with $N+1$ properties. Concretely, we start by considering all subhalo properties when building RFs with one variable, but only keep the best-performing properties as a basis to build the 2-variable RFs, and so forth.

In all cases, we employ the cross-matched catalogues between TNG300-1-Dark and TNG300-1 (c.f. \S\ref{sec:TNG}). We apply a threshold $\mpeak = 3\times 10^{10}\,\hMsun$, corresponding to objects resolved with at least 10 particles in TNG300-3-Dark and TNG300-mimic. To build the RFs, we use the publicly available code \texttt{Scikit-Learn} \citep{scikit}. On a single CPU, the training time varies between 1 and 10 minutes per RF, depending on the number of variables considered. 

\subsubsection{Performance metric}
\label{sec:RFchi}
We split the simulation volume in two halves: one half is used for training the RFs while the other is used for assessing its performance. 

Specifically, for each RF, we compute the predicted SFR in subhalos within the validation half of the TNG300-mimic. We then compute the resulting monopole and quadrupole of the redshift-space correlation function.\footnote{These statistics are computed by measuring the cross-correlation between the validation sample (half of the box) and the full galaxy sample. This selection of samples and clustering function allows us to measure a clustering signal dominated by the validation sample (i.e., galaxies not used in the training process) while reducing the noise on the clustering statistic. This method is repeated twice, changing the side of the box used as the training set. Details about the calculation of the correlation functions can be found in Appendix~\ref{sec:AppClustering}.} These measurements, jointly denoted as $V_{\rm RF}$, are compared with those obtained from the TNG300 $V_{\rm TNG300}$ via

    \begin{equation}
    \label{eq:chi2}
    \chi^2 = (V_{\rm TNG300}-V_{\rm RF})^T C_v^{-1} (V_{\rm TNG300}-V_{\rm RF}),
    \end{equation}

\noindent where $C_v$ is the covariance matrix of the TNG300 estimated with 27 jackknife samples \citep{Zehavi:2002, Norberg:2009} and adding in the diagonal 5\% of the signal to account for other sources of systematic error. We compute the $\chi^2$ associated with a given RF as the average value over two samples $\bar{n} = 10^{-2}$ and $\bar{n} = 10^{-2.5}\,\ihMpcC$ at both $z=0$ and $z=1$. 
	
\subsection{Key subhalo properties}
    \label{sec:RF_results}
    \begin{figure}
		\centering
		\includegraphics[width=0.45\textwidth]{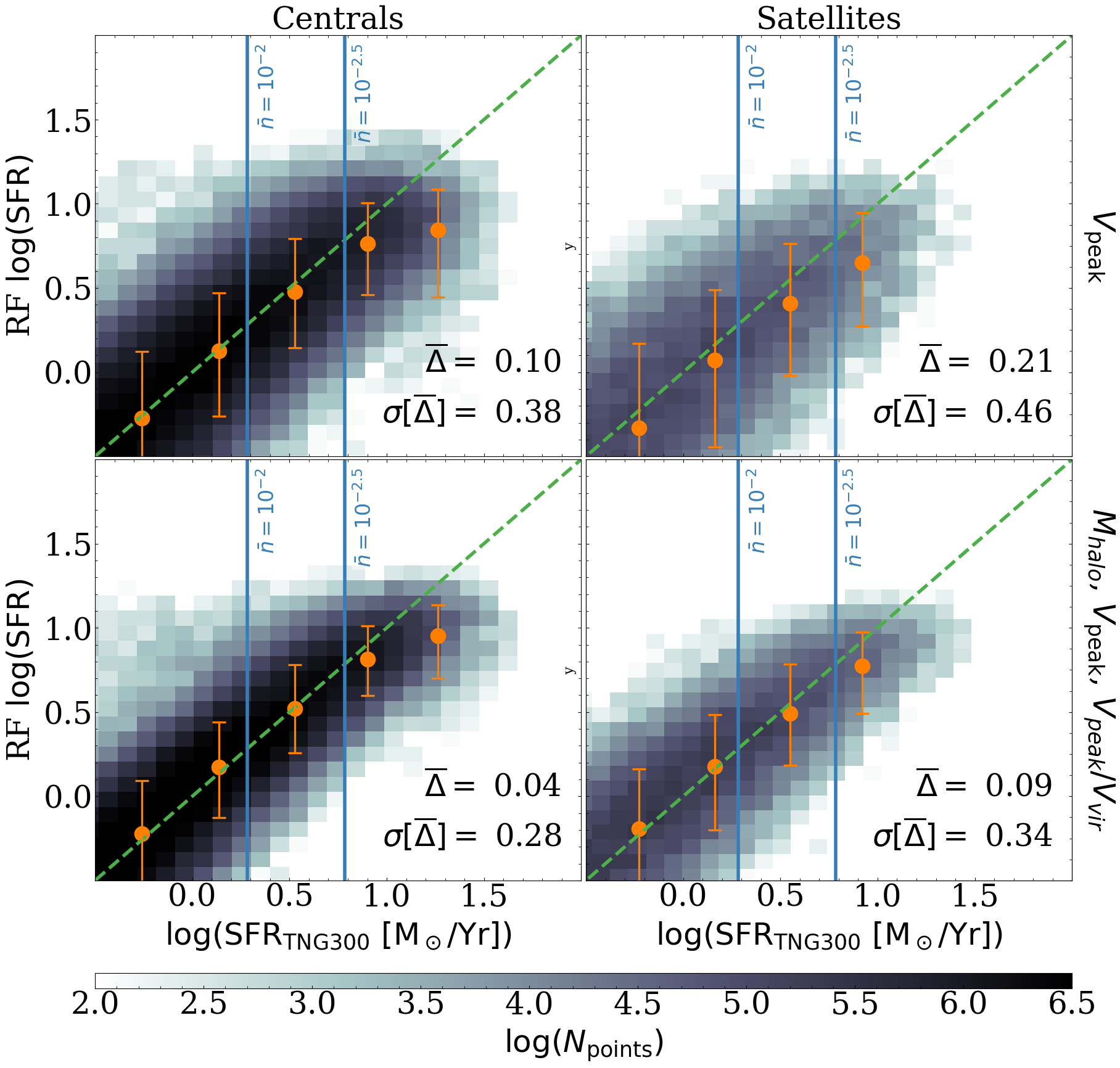}
		\caption{Comparison between the real SFR values (TNG300, x-axis) and the ones predicted by the RF (y-axis) for the validation sample at $z = 1$ using one the best three models with three subhalo properties (bottom) and the model with only $\vpeak$ (top). Blue lines show the SFR values that define SFR-selected number densities of $\bar{n} = 10^{-2}\,\ihMpcC$ and $10^{-2.5}\,\ihMpcC$ for the TNG300 values. Circles with error bars indicate the median, 16$^{th}$ and 84$^{th}$ percentiles, and are only calculated when the number of points in a 0.4 dex bin is larger than 1000. The dashed green line indicates $x = y$. $\overline\Delta$ is the mean deviation between the real and predicted values, and it is calculated in log scale using all the subhalos with log(SFR) $>$ 0, and assuming SFR = $10^{-4}$ M$_\odot$/Yr for quenched galaxies. }		
  \label{fig:RFprediction}
	\end{figure}
 
	\begin{figure*}
		\centering
		\includegraphics[width=0.95\textwidth]{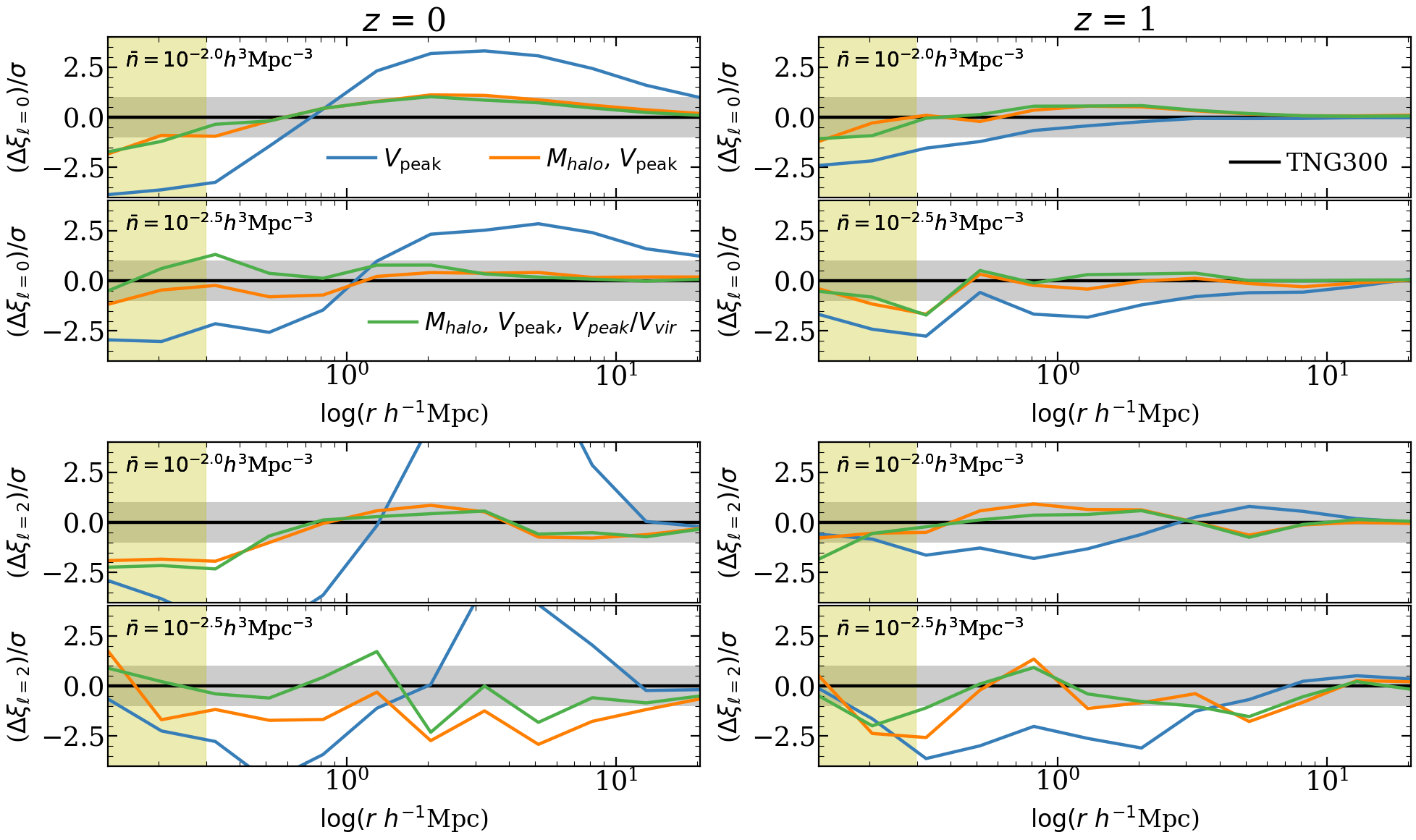}
		\caption{Difference between the predicted and TNG monopole (top block) and quadrupole (bottom block) of the redshift space cross-correlation function between the test half of the box and the whole box. We show three models with different numbers of subhalo properties. We test two redshifts: $z = 0$ (left) and $z = 1$ (right), and two number densities for each statistic ($n = 10^{-2}$$\ihMpcC$, upper panel of each block, and $n = 10^{-2.5}$$\ihMpcC$ lower panel). Grey-shaded intervals mark the $1\sigma$ regions. The yellow-shaded region ($r < 0.316 \hMpc$) is not used to compute the $\chi^2$ (see: Appendix~\ref{app:chi2}). }
		\label{fig:RFresult}
	\end{figure*}

 \begin{figure}
		\centering
		\includegraphics[width=0.45\textwidth]{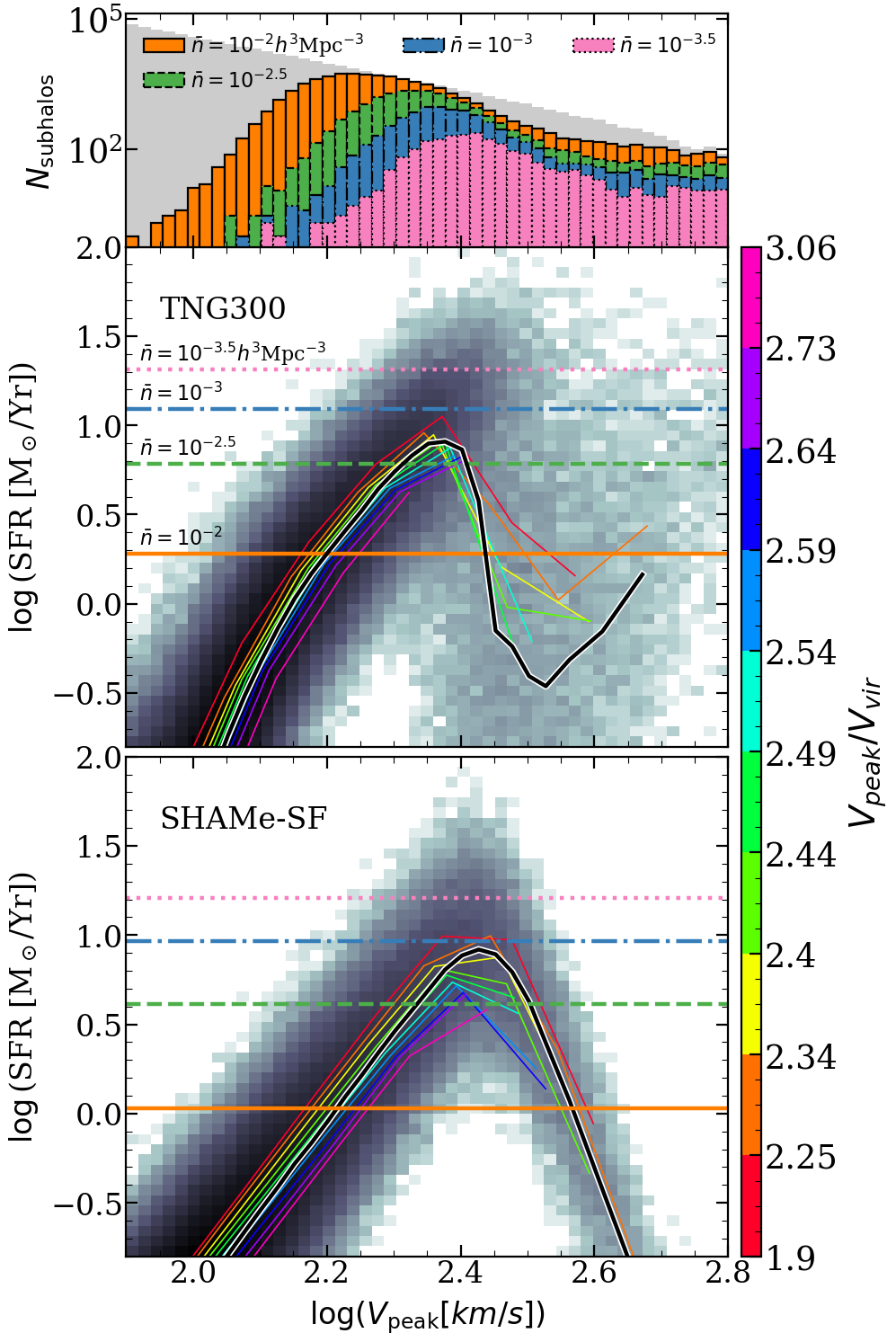}
		\caption{\textbf{Top}: $\vpeak$ distribution for the different number densities analysed in this work: $n = 10^{-2}\ihMpcC$ (orange, solid), $n = 10^{-2.5}\ihMpcC$ (green, dashed), $n = 10^{-3}\ihMpcC$ (blue, dash-dotted), and $n = 10^{-3.5}\ihMpcC$ (pink, dotted). The distribution for all the subhalos is shown in grey.  \textbf{Center and bottom:} $\vpeak$ Vs. SFR distribution for TNG300 (centre) and the SHAMe-SF model (bottom) for central subhalos at $z = 1$. Colour-coding represents the median of $\vpeak/\vvir$-defined intervals. The solid, dotted, dash-dotted and dotted lines mark the values of the SFR that define each number density in the top panel.}
		\label{fig:vpeakSFR}
	\end{figure}

When considering more than one property, the best combinations of two properties were found to be [$\vpeak$, $\mhalo$] followed by [$\mpeak$, $\mhalo$]. It is interesting to note that several previously mentioned models assumed $\dot{M}$ to be a good proxy for SFR. This was also the case in our analysis, but only when considering a single subhalo property -- $\dot{M}$ was not among the best models when the number of subhalo properties was increased. 

When incorporating a third variable, $\vpeak/\vvir$, $\vmax/V_{200,\rm t_{infall}}$ (concentration proxies), $\dot{V}$, and Age emerged as the most informative additional properties when combined with [$\vpeak$, $\mhalo$]. Here we highlight that the combination [$\vpeak, \dot{V}$] is used in \texttt{UniverseMachine} \citep{Behroozi:2019}.  Quantitatively, including the third subhalo property produced only minor improvements in the predicted clustering when compared to models with two properties. Similarly, when including four or more subhalo properties, we find almost no improvement in the clustering predictions.

\subsubsection{Performance of the algorithm}
\label{sec:RFplots}

We now illustrate the performance of our RFs at predicting the SFR of individual subhaloes and the clustering of SFR-selected samples. Specifically, we will display the predictions of our best-performing RF with three properties ($\vpeak$, $\mhalo$ and $\vpeak/\vvir$) and its "root" models with one and two fewer properties.

 Fig.~\ref{fig:RFprediction} displays the SFR for central (left) and satellite (right) subhalos, as measured in the validation set of TNG300 and as predicted by our RFs. Filled symbols with error bars show the median and standard deviations in bins of TNG300 SFRs. Additionally, in the legend we quote the mean logarithmic difference, $\overline{\Delta}$, and mean logarithm error, $\sigma[\overline\Delta]$. We indicate as vertical blue lines the two number density cuts used to quantify the performance of each RF.

We can see that the active subhalos are fairly well predicted by our 1-variable RF using $\vpeak$, performing somewhat better for central compared to satellite subhalos. Adding additional properties reduces the scatter, by about 25\% for centrals and by 35\% for satellites, and halves the bias in the mean predicted SFR. This is an important validation that our RFs are learning the connection between SFR and subhalo properties, also highlighting that the increased number of variables adds non-redundant information.

We note that both RFs underpredict the SFR of the most star-forming galaxies. This might not be surprising considering the influence of the distribution of training values on the final model, where these galaxies are under-represented. RF and other mean-based predictors tend to mispredict the tails of the target distribution (as discussed in \citealt{deSanti:2022}, which includes an analysis of different machine learning methods and the impact of modifications to the input data). To achieve a better modelling of extreme star-forming galaxies, alternative methods capable of capturing the tails of distributions would be necessary. For instance, one possibility is modelling the probability distribution for each value of the input properties, instead of only returning the mean value (e.g., \citealt{Rodrigues:2023}). 

We further explore the accuracy of the RF predictions in Fig.~\ref{fig:RFresult} where we show the clustering at $z = 0$ and $z = 1$ for both of our validation samples $\bar{n} = 10^{-2}$ and $10^{-2.5}\ihMpcC$. In addition to the previous 1 and 3-variable RFs, we include a third RF with 2-variables for comparison. In Appendix~\ref{app:chi2} we expand on the comparison between these and other RFs in terms of their respective $\chi^2$ values (c.f. Eq.~\ref{eq:chi2}).

We can see a significant improvement in the predicted clustering of the 3-variable RF when compared to that built uniquely on $\vpeak$. As we will discuss later, we argue the improvement reflects the RF learning a quenching mechanism and the nature of the scatter for active galaxies. This interpretation is in agreement with the 2-variable RF being statistically consistent with the 3-variable RF. Although not shown here, all the top three 3-variable RFs perform at a similar level of accuracy. 
	
\section{The SHAMe-SF model}
\label{sec:Model}

The previous section showed that the clustering of SFR-selected samples from TNG galaxies appears to be mostly determined by 3 subhalo properties. We now discuss how these properties can be used to build a flexible model that predicts the clustering of star-forming galaxies.

The first aspect to consider is that the relation between DM properties and SFR can be highly nonlinear and difficult to disentangle. For this reason, we start by examining and modelling the SFR for central galaxies, and later explore enhancements required for satellite galaxies. Even if we focus on the subhalo properties selected by the RF, we are open to including additional properties for implementation purposes to avoid complex analytical dependencies.

\subsection{SFR predictions for central galaxies}
    
We start by noting that RFs containing $\vpeak$ performed slightly better than those using $\mpeak$ as the primary property. This might be because $\vpeak$ is more sensitive to inner regions of DM haloes compared to $\mpeak$. Consequently, we will select $\vpeak$ to build the SHAMe-SF model. Nonetheless, we expect that a model based on $\mpeak$ would lead to similar results.     
    
The middle panel of Fig.~\ref{fig:vpeakSFR} displays the distribution of [$\vpeak$, SFR] values for central subhalos with SFR $ > 10^{-4}$ $\rm M_\odot/Yr$ in the TNG300 at $z=1$. The total number of subhaloes as a function of $\vpeak$ in various samples, as indicated by the legend, is shown in the top panel.

Firstly, we note that the mean $\vpeak$-SFR relation, indicated by a thick black line, is non-monotonic. Qualitatively, this relation reflects that SFR increases with the size of the DM host due to the increment of the available cold gas. This relation breaks at the point where the central supermassive black hole and its accretion rate are large enough for AGN feedback to become effective, which quenches the galaxy decreasing its SFR. For even larger host haloes, feedback in TNG300 seems to become incapable of counteracting the effect of star formation and SFR increases again for subhalos with $\log(\vpeak$ $[km/s])>2.5$. 

Neglecting the SFR increase at high $\vpeak$, the overall $\vpeak$-SFR can be described by a broken power-law: 

\begin{equation}
		{\rm SFR}|_{({\vpeak})} \propto \frac{1}{\left( \frac{\vpeak}{V_{1}}\right)^{-\beta} + \left( \frac{\vpeak}{V_{1}}\right)^{\gamma} },
    \label{eq:SFR}
\end{equation}
	
\noindent where $\beta$ and $\gamma$ are the values of the slopes around $V_1$, the value of $\vpeak$ at the turnover of the relation. 

Secondly, we appreciate a large scatter ($\sim0.5$dex) in the $\vpeak$-SFR distribution. After considering all the properties identified RF algorithm in addition to $\vpeak$, we find that $\vpeak/\vvir$ strongly correlates with deviations from the median $\vpeak$-SFR relation. We note that other properties that also could parametrize the deviations from the main relation were $\dot{V}$ and subhalo age, but their effect was more subtle. We performed the same tests analysing the $\mpeak$-SFR, but the combination of $\vpeak$ and $\vpeak/\vvir$ returned the best description of the SFR distribution.

In Fig.~\ref{fig:vpeakSFR} we also show the mean relations in bins of $\vpeak/\vvir$, as labelled by the vertical colour bar. In the case of NFW profiles, $\vpeak/\vvir$ can be directly linked to the concentration parameter. Therefore, we can see that less concentrated subhalos host galaxies with higher SFRs, which we interpret as galaxies populating more concentrated subhalos tend to form earlier, running out of gas and thereby exhibiting lower SFR \citep{Lin:2020}.   

To implement the previously discussed dependencies into our empirical model we proceed as follows:

    \begin{enumerate}
    \item We compute the SFR as a function of $\vpeak$ using, using three free parameters, $\beta$, $\gamma$, $V_1$ (c.f. Eq.~\ref{eq:SFR})
    \item We add a random scatter given by a Gaussian with width $\sigma$ (another free parameter) constant in log(SFR).
    \item For a fixed value of $\vpeak$, we semi-sort the scatter using $\vpeak$/$\vvir$ following the method described in \cite{Contreras:2021bias}, which yields a distribution ordered with some random noise (given by the free parameter $f_k$). Central and satellite subhalos are sorted separately to keep the satellite fraction constant. We refer to this as ${\rm SFR}|_{({\vpeak,\vvir})})$.
    \end{enumerate}
    
 We show the predicted distribution of SFRs and their dependence with $\vpeak/\vvir$ on the lower panel Fig.~\ref{fig:vpeakSFR} (using a preliminary set of parameters). We see a remarkable similarity with the results from TNG300, with, perhaps, the exception of subhalos with $\log \vpeak > 2.6$. As mentioned earlier, this feature might originate by inefficient AGN feedback in TNG300. Note, however, that this feature is absent in other models (see \citealt{C15} for some SAM examples). These differences also appear when comparing models to observations, and arise from the implementation of quenching processes and environmental effects \citep{Popesso:2015}. However, those high-$\vpeak$ subhaloes are very rare and thus subdominant for clustering predictions. Therefore, we have not adopted a more complex model by default, although SHAMe-SF can be trivially extended in the future to incorporate such a feature. 
   
\subsection{SFR predictions for satellite galaxies}
    
  As discussed earlier, we expect the modelling of SFR in satellite galaxies to be more complex than that of centrals. When a galaxy becomes a satellite, additional physics is at play: tidal forces start stripping its cold gas, which finally quenches the galaxy (generally after the first pericenter passage, \citealt{Orsi:2018}). 
    
\begin{figure}
		\centering
		\includegraphics[width=0.45\textwidth]{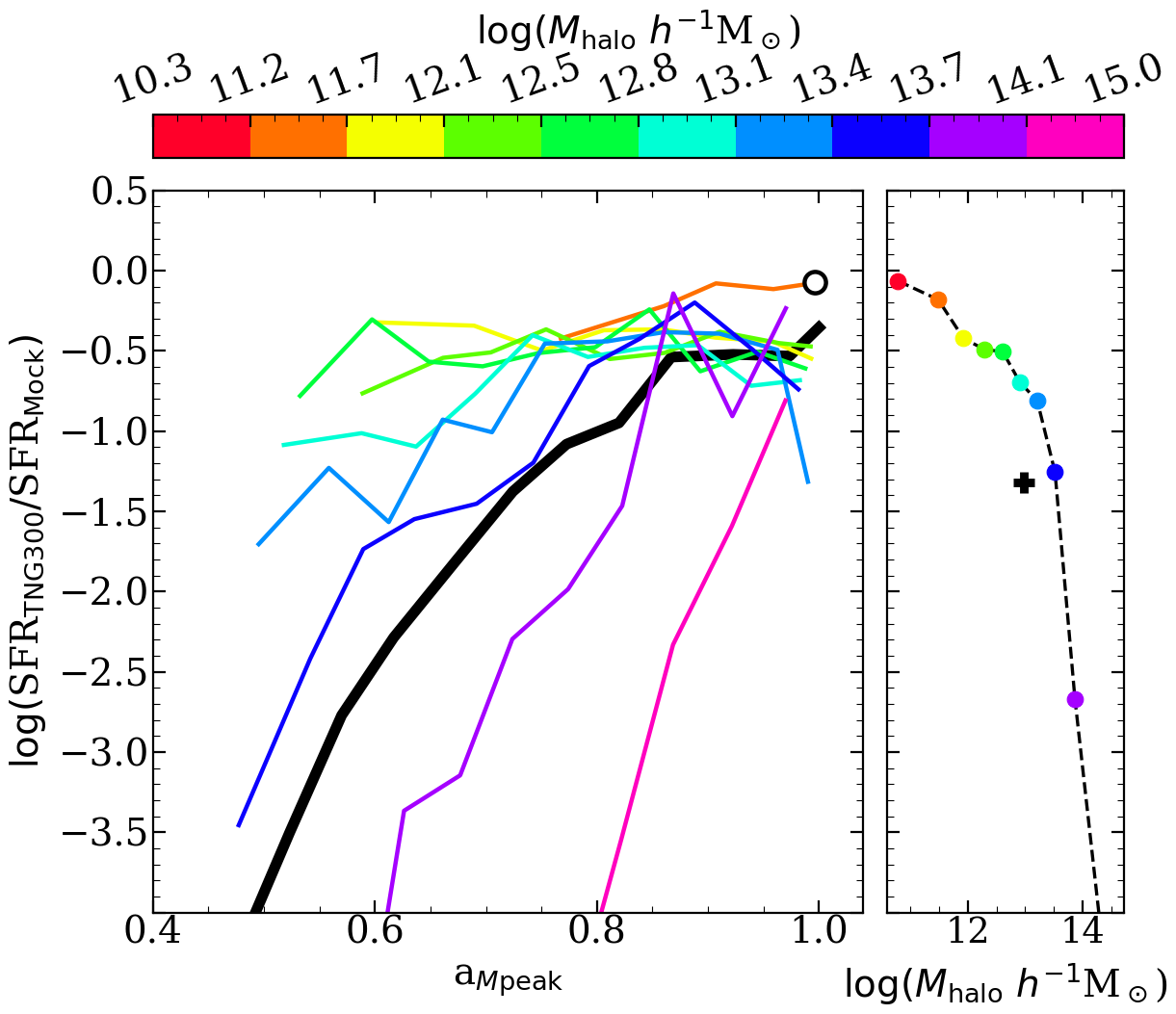}
		\caption{For satellites, mean difference (in log scale) between the TNG300 values and the predicted ones using only Eq.~\ref{eq:SFR} and the semi sorted scatter for a subsample of galaxies on a fixed [$\vpeak, \vpeak/\vvir$] interval. \textbf{Right:} for different $\mhalo$ values (colour coding). \textbf{Left:} for the same $\mhalo$ intervals, plotted as a function of $\tmpeak$ (in scale factor units). The black thick line shows the behaviour when the whole sample is considered. The circle represents the mean value for central galaxies with $\tmpeak \sim 1$. }
		\label{fig:satdifference}
	\end{figure}

To model this additional complexity, we first explore if the difference in SFR between our model so far and TNG300 correlates with the host halo mass, the second property selected by the RF. This is shown in the right panel of Fig.~\ref{fig:satdifference} where we display the average ratio between the SFR in our model and in TNG300. We show results for satellite subhalos in bins of $\mhalo$, as labelled in the top colour bar. Note we plot results for a fixed range in $\vpeak$ and $\vpeak/\vvir$ to avoid dependences on the relation already modelled. 

We clearly see that our model predicts roughly the correct SFR of satellites host by halos of $10^{11}\hMsun$ but significantly underpredicts it in larger haloes. For instance, at $10^{14}\hMsun$ haloes, SFRs in the TNG are about a factor of $10$ smaller than in our model. This is not surprising since more massive haloes quench galaxies faster, mainly by to their inter-cluster medium (ICM) striping the cold gas of the galaxies that form stars \citep{ChavesMontero:2016} -- a process absent in our model so far.

To further explore the modelling of quenching, we plot in the left panel of Fig.~\ref{fig:satdifference} the ratio of the model and TNG300 SFR as a function of $\apeak$. We show results in bins of host halo mass at $z = 0$, where quenching processes are more statistically significant \cite{Donnari:2020}. However, similar trends appear at $z = 1$. 

We can clearly see that galaxies with $\apeak \sim 1$ (i.e. those recently accreted) have similar SFRs as those of centrals. Smaller values of $\apeak$ typically consist of more quenched galaxies. This is consistent with environmental processes in clusters, where a higher accretion redshift results in more efficient stripping. Although not shown here, other galaxy samples and bins in $\vpeak/\vvir$ display consistent results.

We find that we can parametrise the decrement of SFR as a power-law, with a different slope depending on the host halo mass:

    \begin{equation}
		\log(\rm SFR) =  \log_({\rm SFR|_{({\vpeak,\vvir})})} - \alpha_{\mhalo}\times (a_z - \tmpeak),
    \label{eqn:quenching}
	\end{equation}
	
\noindent where a$_{z}$ is the value of the expansion factor and $\alpha_{\mhalo}$ is given by:
	
	\begin{equation}
		\alpha_{\mhalo} =
		\begin{cases}
			\alpha_0 \left[\log\left( \frac{\mhalo}{M_{\rm{crit}}}\right)\right]^{\alpha_{\exp}} &\text{if $\mhalo > M_{\rm{crit}}$}\\
			0                                                                                         &\text{if $\mhalo \leq M_{\rm{ crit}}$}\\
		\end{cases}   
        ,
        \label{eq:qdefinition}
	\end{equation}
	
\noindent where $\alpha_0$, $\alpha_{\rm exp}$ and $M_{\rm{ crit}}$ are free parameters. We force $\alpha_{\mhalo}$ > 0 to avoid the enhancement of the SFR for satellites on lower mass halos. We perform the same test for central subhalos, finding the same behaviour in the (rare) cases of subhalos whose peak time was in the past.

We highlight that $\tmpeak$ was not on the list of best properties suggested by our RF analyses. However, we recall that our performance metric was given by general clustering statistics for which the mass-dependent quenching of satellite galaxy might have a minor effect, specially when considering the statistical noise in the TNG300 catalogues (we compare a RF with this property in the Appendix~\ref{app:chi2}). Given the clear trends shown in Fig.~\ref{fig:satdifference}, we decided to include it in the model foreseeing additional applications to observable more sensitive to the intra-halo quenching.

Summarising, SHAMe-SF models the SFR of galaxies as a function of $\vpeak$ (Eq.~\ref{eq:SFR}). Deviations from this mean relations are assumed to partially correlate with $\vpeak/\vvir$. The SFR in satellites is further modulated as a function of the host halo mass and time since accretion (Eqs~\ref{eqn:quenching} and~\ref{eq:qdefinition}).

\section{Redshift space clustering}
\label{sec:clusteringall}

In this section, we examine the ability of SHAMe-SF to describe the projected and redshift-space correlation function of SFR-selected samples. First, we present an emulator for SHAMe-SF. Then, we describe our fitting procedure and show the results for $z = 1$ for TNG300, SAM, and DESI-like samples. We finalise with a discussion on the impact of the free parameters of our model. 
 
\subsection{SHAMe-SF emulator}	
 \label{sec:clusmethod}

\begin{figure*}
    		\centering
            \includegraphics[width=0.31\textwidth]{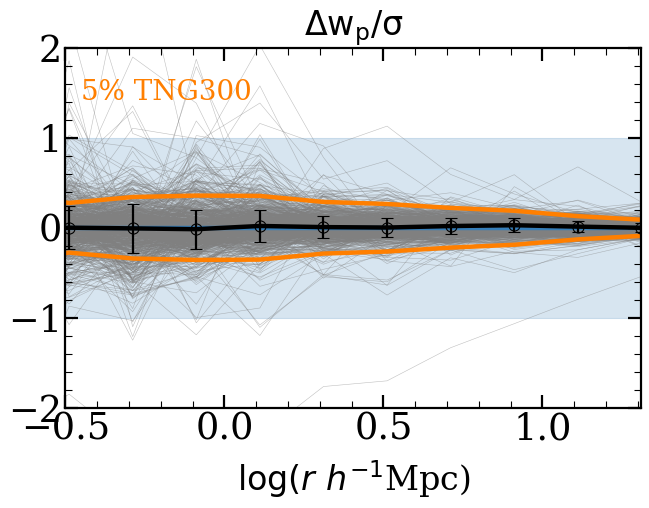}
            \includegraphics[width=0.31\textwidth]{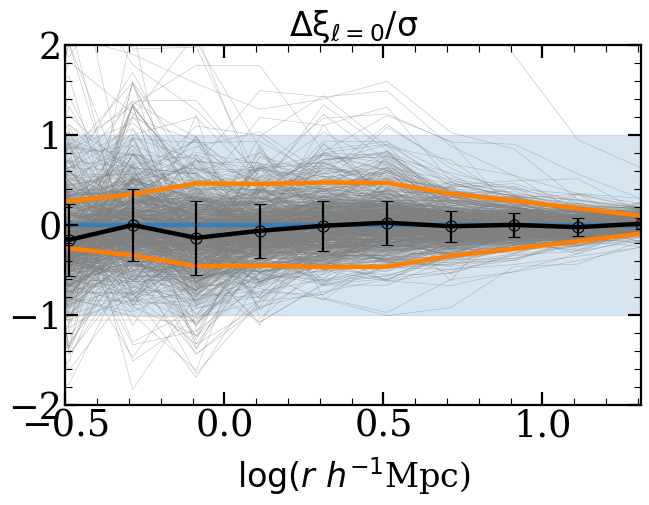}
            \includegraphics[width=0.31\textwidth]{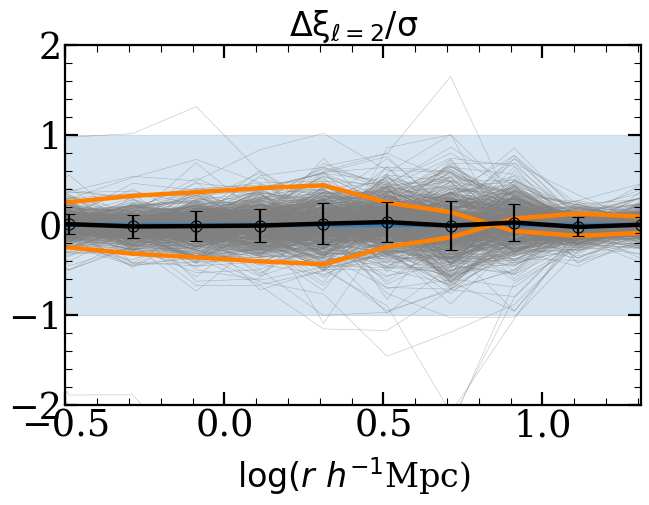}
    		\caption{Difference between the emulator prediction and the real mock weighted by the scaled TNG300 error for the highest number density ($\bar{n} = 10^{-3}\,\ihMpcC$) at $z = 1$. Error bars mark the 68\% intervals of the distribution of the emulator predictions. We plot 800 randomly selected lines in grey. We include in orange the contribution to the total uncertainty ($\sigma$) of the 5\% of the TNG300 signal. The total uncertainty (including the JN samples) is marked using the blue shaded region for visual aid.}
    		\label{fig:emulator}
\end{figure*}

Creating one SHAMe-SF mock on the TNG300-mimic (Section~\ref{sec:TNGmimic}) takes around 1 CPU second, needing an additional 2 seconds to compute the clustering statistics (for details see Appendix~\ref{sec:AppClustering}). While reasonably fast, this becomes computationally expensive for Monte-Carlo analyses. 

Following \cite{Arico:2020,  Arico:2021, Angulo:2021,  Zennaro:2021, Pellejero:2023,  C2023:lensing}, we speed up the evaluation of the model by constructing Neural Network emulators. Specifically, we use the same architecture as \cite{C2023:lensing}: two (three) fully connected hidden layers with 200 neurons each. Alternative configurations yield comparable performances. To build the emulator we use the Keras front-end of the Tensor-flow library. We chose the Adam optimisation algorithm with a learning rate of 0.001 with a mean squared error loss function. We leave 10\% of the data set for validation.

We build 18 different emulators, one for each clustering statistic (projected correlation function, monopole and quadrupole), number density ($\bar{n} = \{10^{-2},10^{-3}, 10^{-2.5}\}\,[\ihMpcC]$) and redshift ($z = 0$ and $1$). We train our emulators by measuring the clustering for 50,000 sets of SHAMe-SF parameters generated by a Latin-Hypercube algorithm over the range:

\begin{eqnarray}
        \label{eq:par_range}
        \beta & \in & [0,20] \nonumber \\
        \gamma & \in & [-15,25] \nonumber \\
        V_1 & \in & [10^{1.8},10^{3.2}] \rm{ (km/s)} \nonumber \\
        \sigma & \in & [0,2]\\
        f_k & \in & [-1,1] \nonumber \\
        \alpha_0 & \in & [0,5] \nonumber\\
        \alpha_{\rm exp} & \in & [-5,5] \nonumber \\
        \log(M_{\rm crit}) & \in & [9,15] \ (\log(M_\odot/h)), \nonumber
        \label{eq:parrange}
    \end{eqnarray}
    
\noindent For each parameter set, we average the predictions of 4 mocks generated by different random realisations of the semi-sorting step. We refer to Section~\ref{sec:Model} for further details on this and the rest of model parameters.        
    
The accuracy of our emulators is shown in Fig.~\ref{fig:emulator} which compares its predictions with measurements in the validation set for the densest galaxy sample at $z=1$. Note we display the difference in units of the statistical uncertainty, $\sigma_{jn}$, as estimated by 27 jackknife samples of the TNG300 plus a 5\% of the signal in the diagonal (shown by the orange line). Note that we renormalise $\sigma_{jn}$ by the ratio between the SHAMe-SF and the TNG300 clustering to account for the differences between the amplitude of the predictions. The emulator uncertainty is typically smaller than the statistical precision of TNG300 in all cases, which also holds for all the other samples not shown.
 
\subsection{Parameter fitting}

    Once we have built the emulators, we employ a Markov Chain Monte-Carlo (MCMC) algorithm to find the SHAMe-SF parameters that best describe the clustering of various samples. We will jointly consider the projected correlation function, and the monopole and quadrupole of the redshift-space correlation function, over the range $r = [0.6, 30]\,\hMpc$. We assume the likelihood of model parameters to be given by a multivariate Gaussian, $\log \mathcal{L} = -\chi^2/2$, where $\chi^2$ is computed using  Eq.~\ref{eq:chi2}. For simplicity, we employ the same covariance matrix described previously discarding off-diagonal terms. We have checked our results were insensitive to using the full covariance matrix. 
    
    We sample the posterior distribution function using an ensemble MCMC algorithm as implemented in \texttt{emcee} \citep{emcee}. We use 1000 chains with 40000 steps each, discarding the first 10000 as a burn-in phase. This somewhat unusual configuration is efficient for our emulator since the computational cost of evaluating a large number of points simultaneously is low (100,000 samples are evaluated in 2 seconds). We run an MCMC for each number density, which takes 90 CPU minutes, and utilise flat priors on model parameters across the full range where the emulator was trained.
    
    Given the small volume of TNG300, the measured clustering of sparse samples has considerable stochastic noise, which complicates the emulator training. For this reason, in these cases we employ a particle swarm optimisation \citep[PSO,][]{PSO} \footnote{A PSO algorithm uses a group of ``particles'' to explore the parameter hyperspace. Their wandering directions are defined by a combination of their individual position, the average position of the swarm, and an inertia component. The positions are given by a target function ($\chi^2$ in our case).} We use the implementation from \cite{PSOBACCO} \footnote{\url{https://github.com/hantke/pso_bacco}} to find the best-fitting SHAMe-SF parameters for the sample with $\bar{n} = 10^{-3.5}\ihMpcC$, for samples at $z=1.5$, and for our DESI-like ELG sample.

 \subsection{Redshift space galaxy clustering predictions}
 \label{sec:clustering}
 
       We now present the SHAMe-SF models that best-fit the clustering of our mock galaxy samples.
      
\subsubsection{TNG300}  
\begin{figure}
		\centering
		\includegraphics[width=0.43\textwidth]{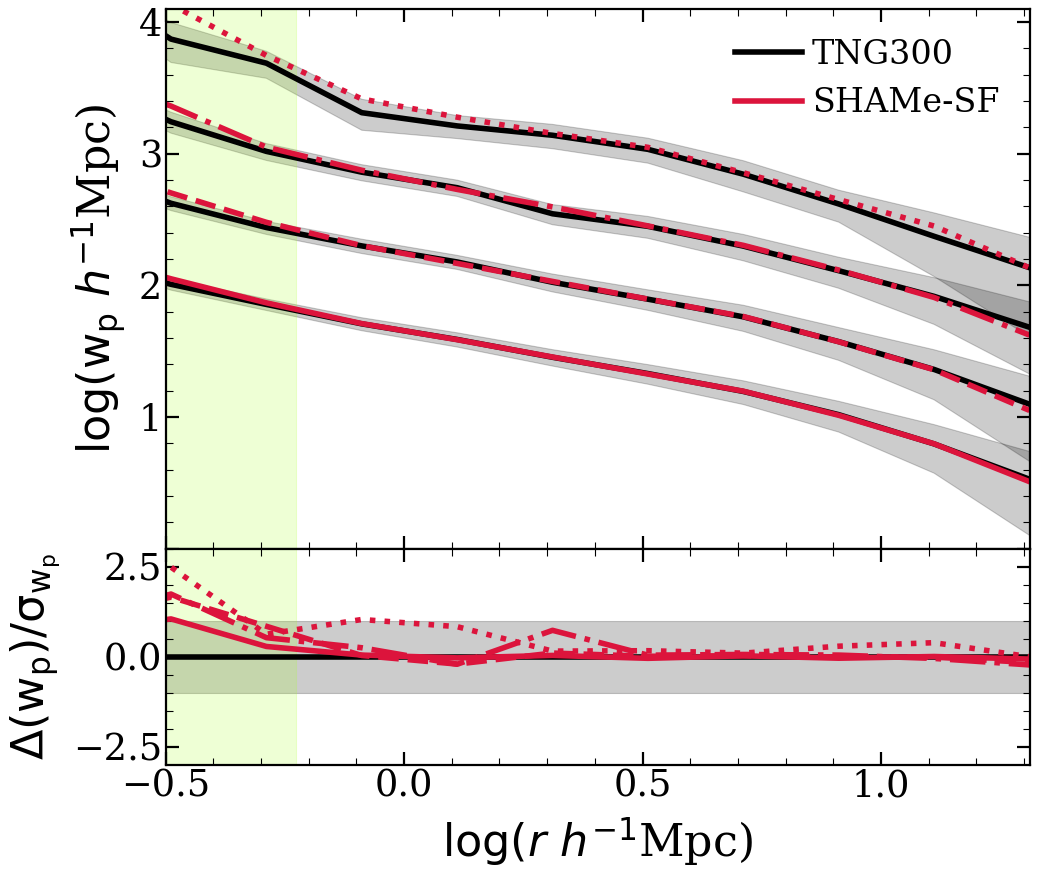}

        \includegraphics[width=0.43\textwidth]{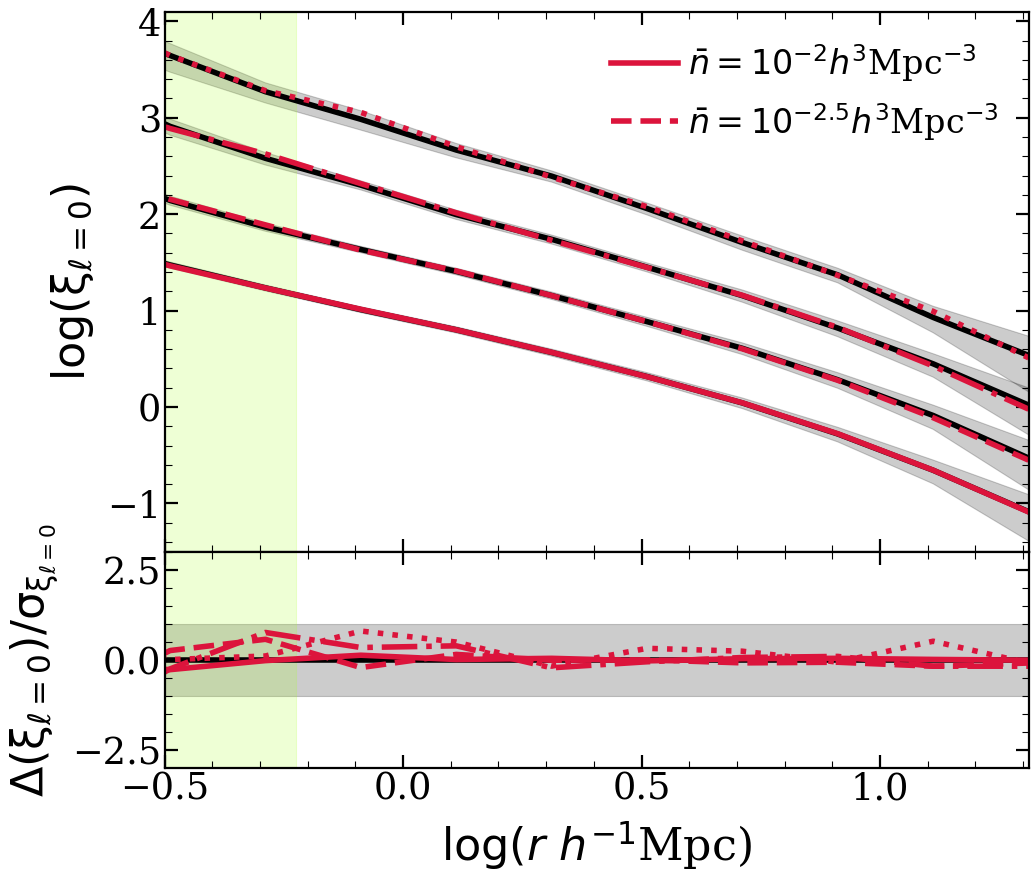}

        \includegraphics[width=0.43\textwidth]{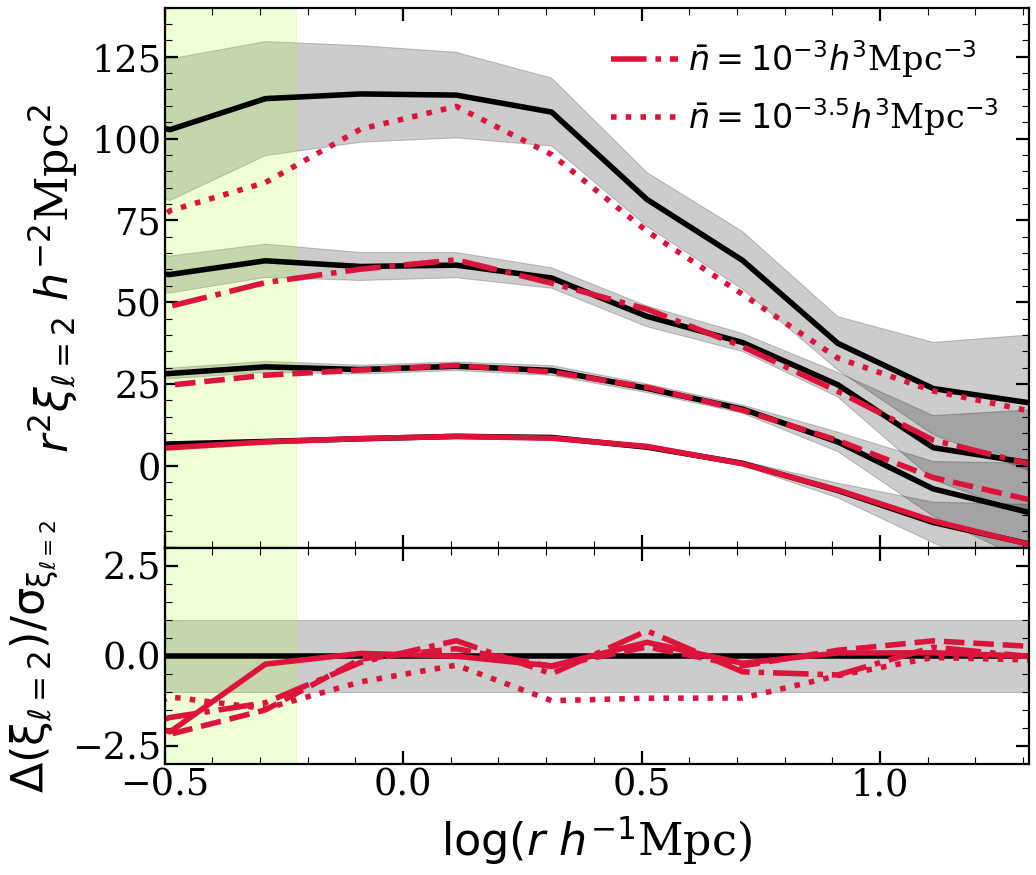}
  
  \caption{Projected correlation function ($w_p$), monopole and quadrupole of the galaxies of the TNG300 simulation (black, $1\sigma$ interval in grey) and the model (red) for a SFR-selected subsample with number densities $n = 10^{-2}$ (straight), $10^{-2.5}$ (dashed), $10^{-3}$ (dash-dotted) and $n = 10^{-3.5}$ (dotted) $\ihMpcC$  at $z = 1$. The three lower number densities have been shifted on the upper panel along the y-axis for visualisation purposes. The lower panel shows the relative difference between the fits and the model normalised by the uncertainty on the measurement on TNG300, with the grey shaded region indicating $1\sigma$. The green-shaded region (${\rm r < 0.6\  Mpc}/h$) indicates the region not considered on the parameter fitting.}
		\label{fig:clusteringz1}
	\end{figure}

 Results for $z = 1$ are shown in Fig.~\ref{fig:clusteringz1} for the projected correlation function (top), monopole (middle) and quadrupole (bottom). We display the results for different number densities using line styles, as indicated by the legend. Higher number densities have been displaced on the y-axis for visualisation purposes. Black line and grey shaded regions represent the TNG300 measurements and its uncertainty, respectively. The lower panel shows the difference between the TNG300 prediction and our best fit, in units of the uncertainty. 
    
    Our model can describe the TNG300 clustering prediction within $1\sigma$ for all statistics and number densities. This is remarkable, considering the difference in resolution between TNG300 and TNG300-mimic, and that the latter is a gravity-only simulation. Note that the accuracy is better for higher number densities where cosmic variance (which is suppressed as TNG300 and its TNG300-mimic have identical initial conditions) dominates the error budget compared to sparser samples where Poisson noise would be more important.  It is interesting to note that we find fits of similar quality as those using RF predictions (c.f. Fig.~\ref{fig:RFresult}), despite modelling simplifications in SHAMe-SF. This implies that our model is effectively capturing all the information about SFR contained in the properties of DM structures. 
    
As mentioned earlier, we only fit scales above $0.6\,\hMpc$ indicated by the green shaded region). This was motivated by \cite{Contreras:2021shame}, who found that SHAMe performed accurately and robustly above those separations. In our cases, we see that the best-fit model is still a good description of the data down to 300 $h^{-1}$kpc, which offers the possibility of describing even smaller scales than in the original SHAMe model. This is also the case for other redshifts and samples (see  Appendix~\ref{app:clusteringmodel}). 

Although we do not show a direct comparison here, we note that, in general, SHAMe-SF performs better than the original SHAMe, delivering more accurate fits, especially for number densities below $10^{-2.5}\,\ihMpcC$.

 \label{sec:SAMcluster}
    \begin{figure*}
    		\centering
    		\includegraphics[width=0.33\textwidth]{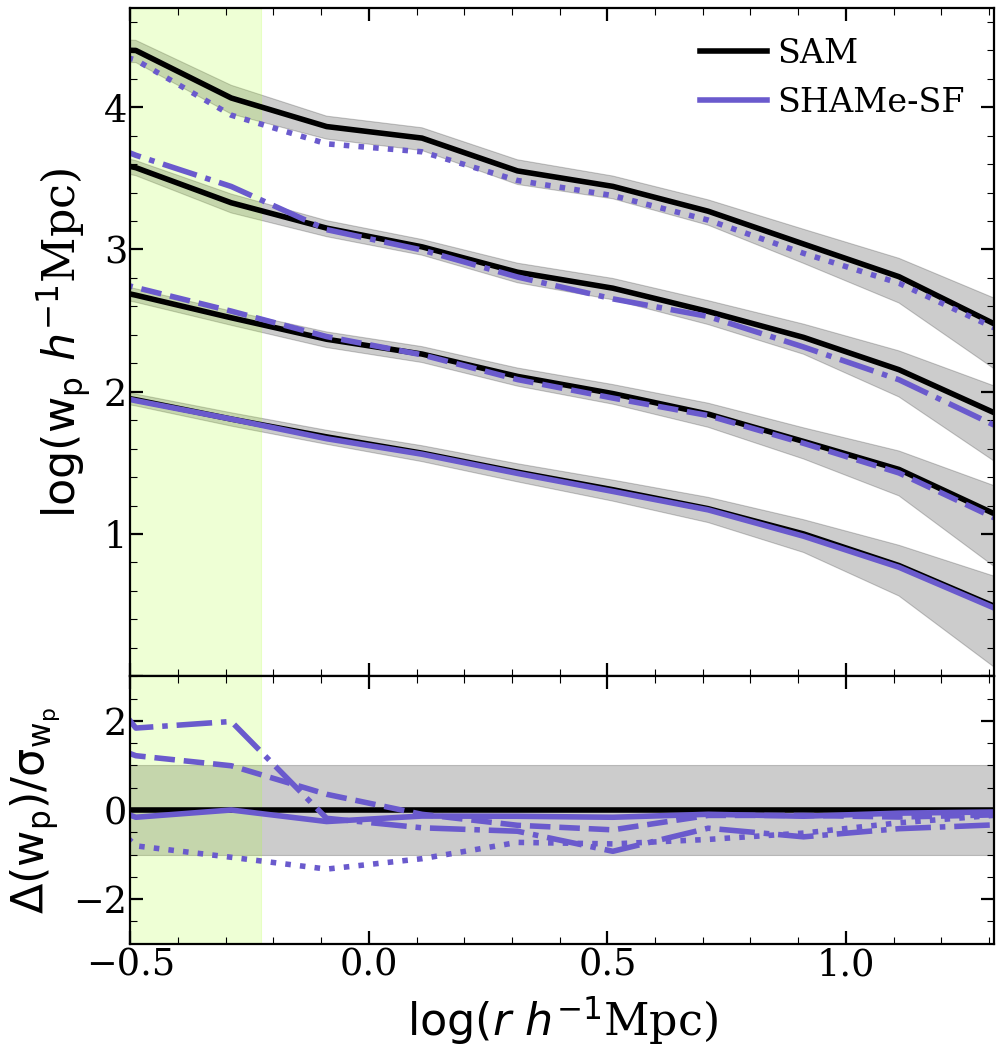}
            \includegraphics[width=0.33\textwidth]{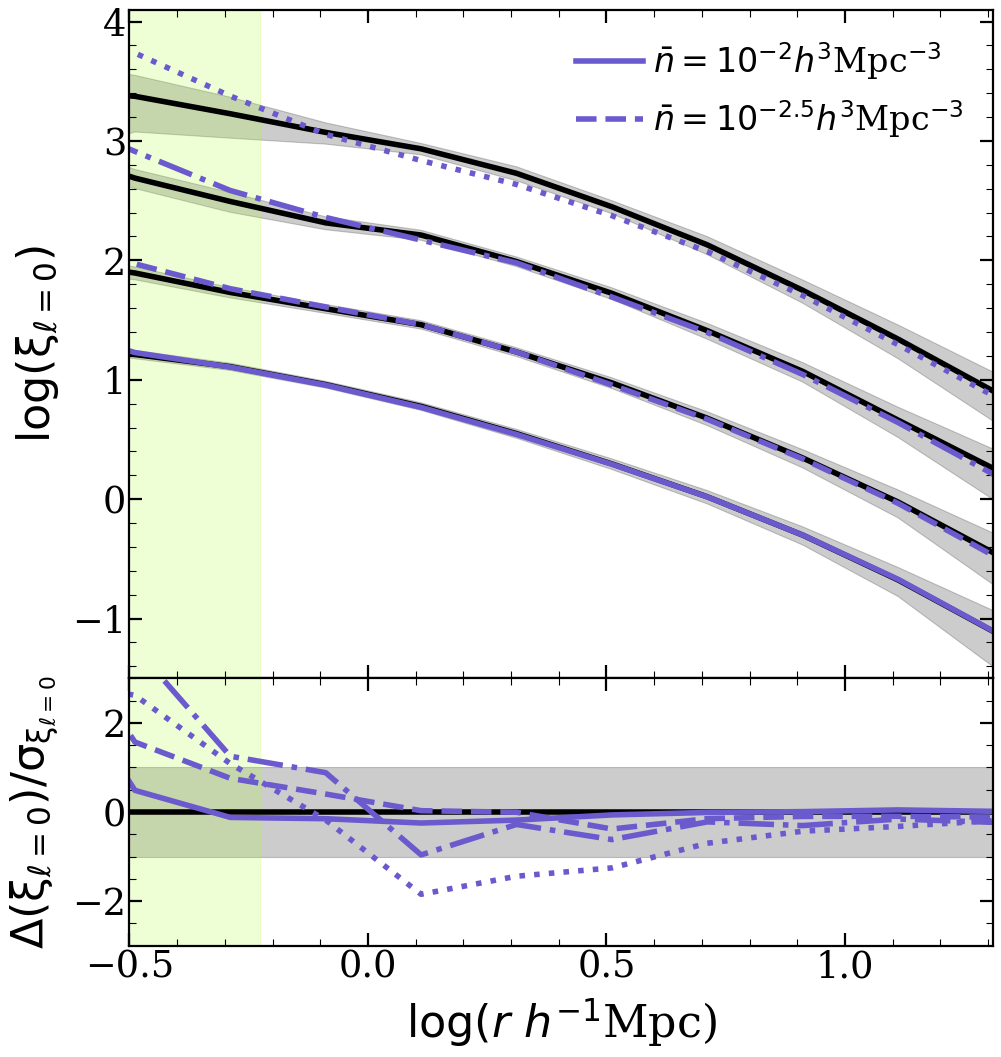}
            \includegraphics[width=0.33\textwidth]{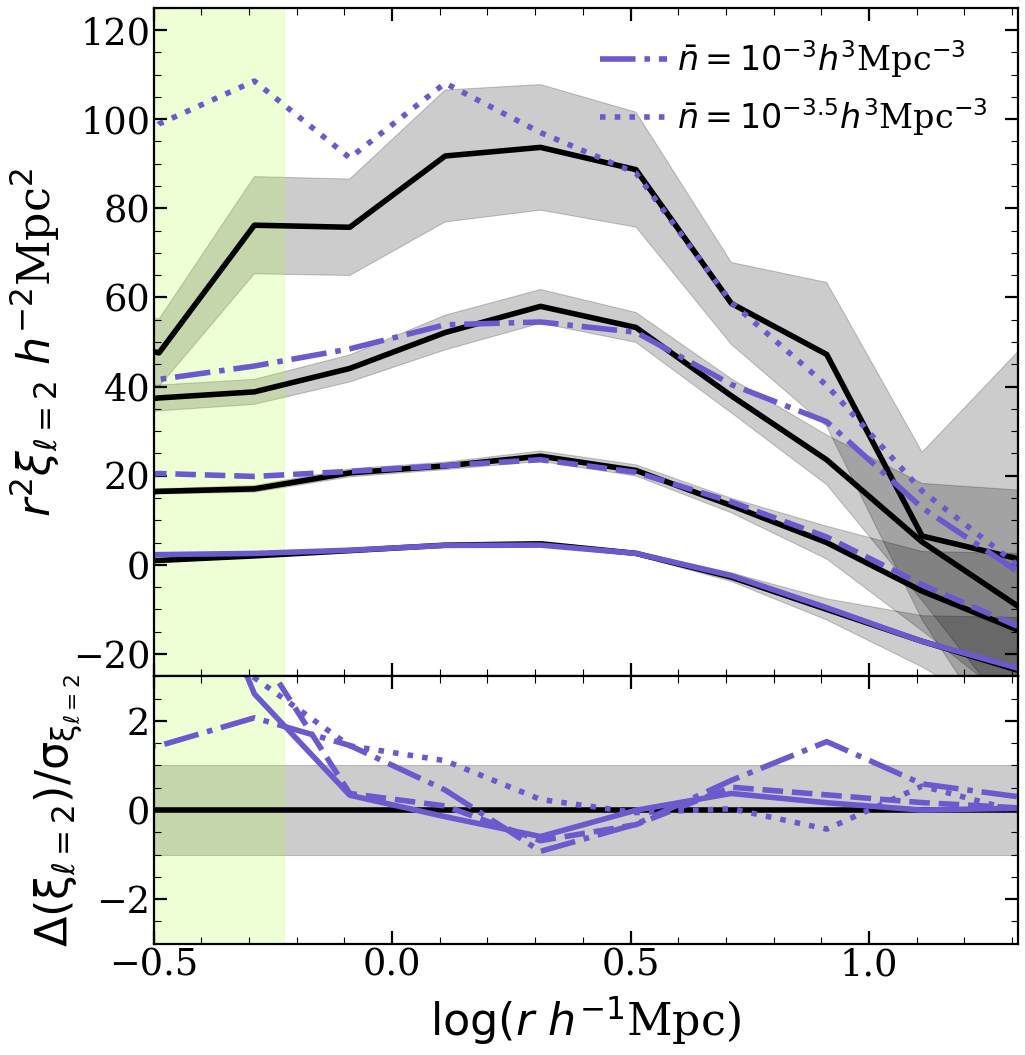}
    		\caption{Same as Figure~\ref{fig:clusteringz1}, but for galaxies from \texttt{LGalaxies} semi-analytical model (black, $1\sigma$ interval in grey) instead of the TNG300 simulation. Mock fit is shown in blue. }   		
      \label{fig:SAMz1}
	\end{figure*}

\subsubsection{SAM}   
    Before carrying out MCMC fits with SAM samples, we confirmed that the basis relations between subhalo properties and SFR discussed in Section~\ref{sec:Model} follow similar trends as in the TNG300. We discuss these relations with further detail in Appendix~\ref{sec:AppLgalaxies}. The main difference between SAM and TNG300 predictions occurs in subhalos with large $\vpeak$ values, where quenching is not as evident as in TNG300: SAM galaxies show SFRs that, on average, monotonically increase with $\vpeak$. We expect, however, that the flexibility of our model is able to capture this feature (by considering negative values of $ \gamma$, the 2nd slope of the SFR-$\vpeak$ relation). 
    
Therefore, we expect SHAMe-SF to provide a reasonable description of the clustering in our SAM samples. This is indeed the case, as shown in Fig.~\ref{fig:SAMz1} which, analogous to the case of TNG300, compares the best-fit SHAMe-SF with the clustering of samples at $z=1$. As in the case of TNG300, we appreciate that our model is statistically consistent with the data over the full range of statistics and scales included in the fit. 
	
\subsubsection{DESI-like ELG galaxies}
 \label{sec:ELGclustering}
    \begin{figure*}
    		\centering
    		\includegraphics[width=0.33\textwidth]{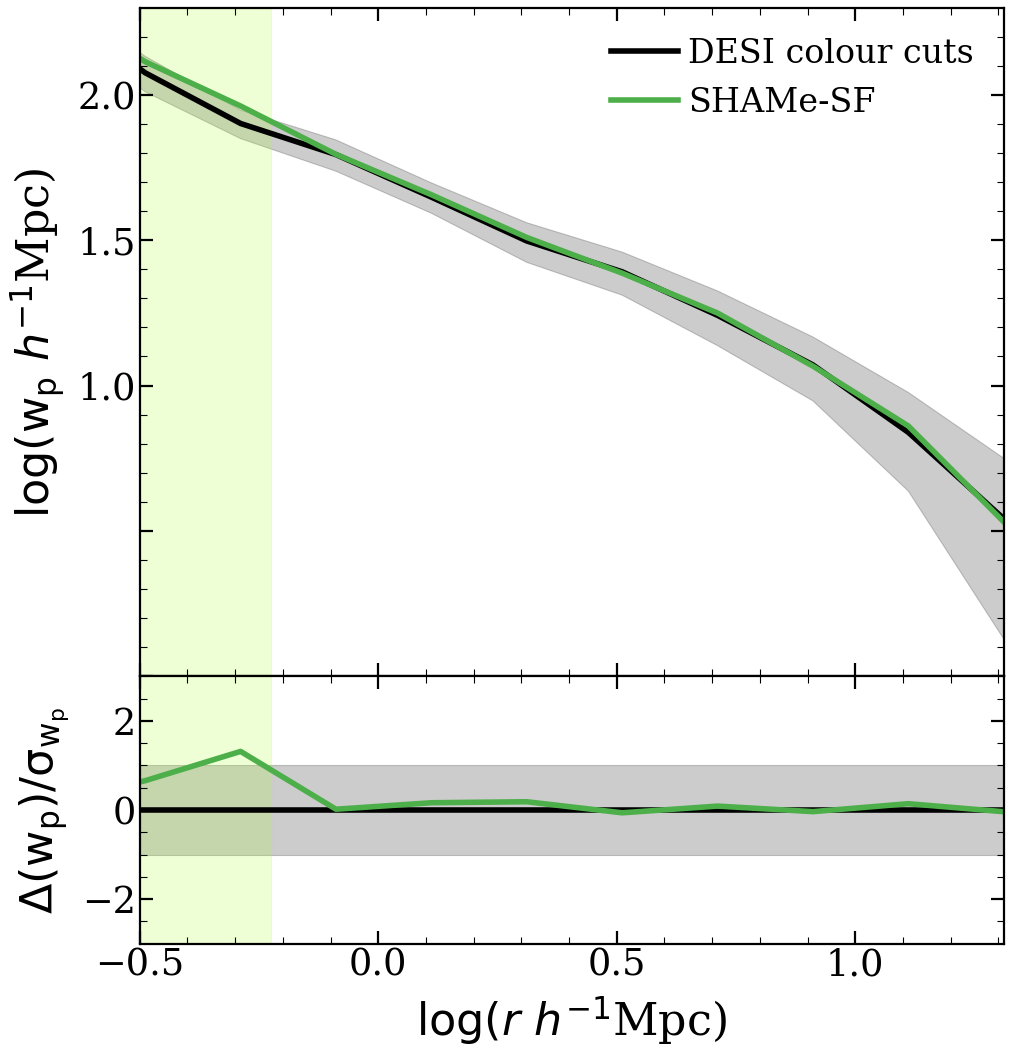}
            \includegraphics[width=0.33\textwidth]{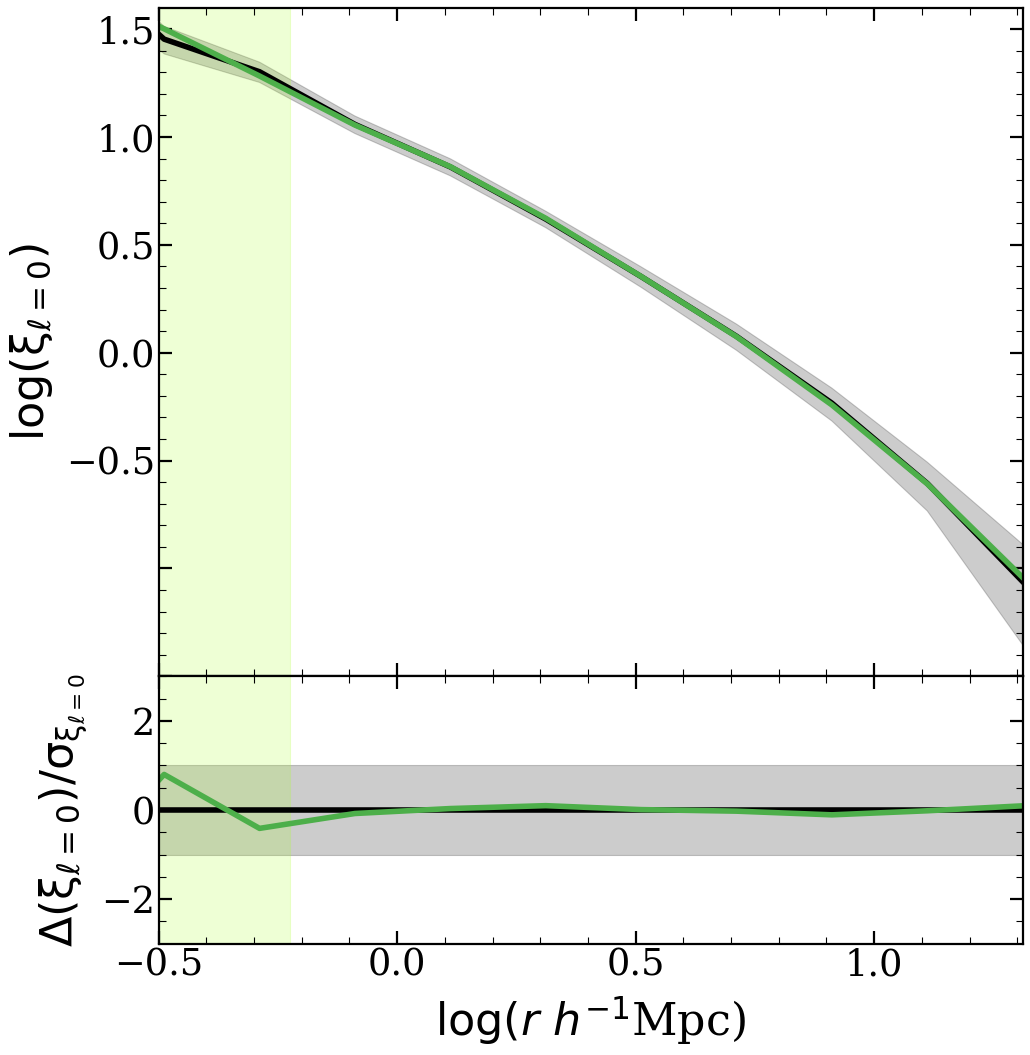}
            \includegraphics[width=0.33\textwidth]{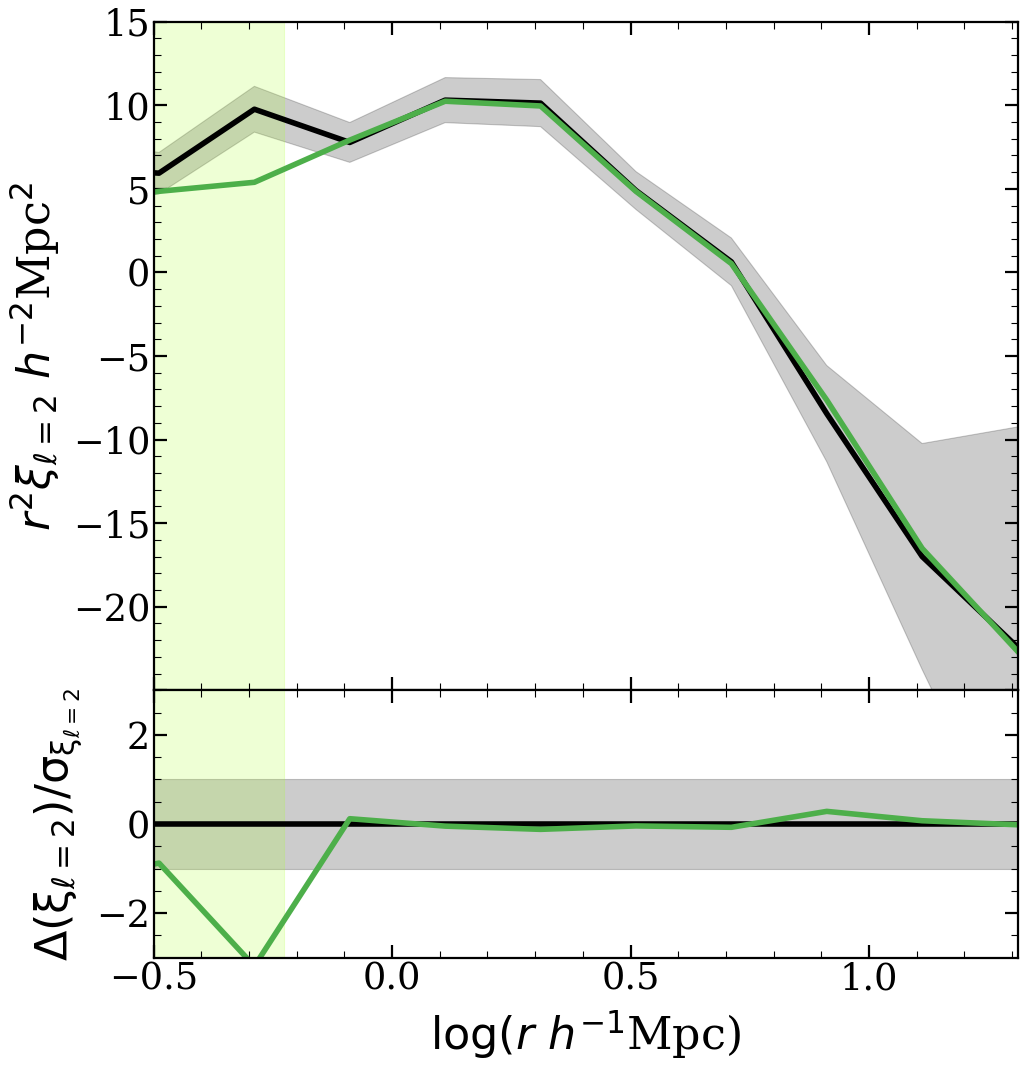}
    		\caption{Same as Figure~\ref{fig:clusteringz1}, but for galaxies applying the colour cuts from DESI (see Table~\ref{tab:ELG}) to TNG300 galaxies at $z = 1$ ($\bar{n} \sim 0.002\,\ihMpcC$). Mock fit is shown in green.}
    		\label{fig:ELG}
	\end{figure*}

    Since the goal of our model is to reproduce the clustering of ELG-like samples, we also fit the clustering of a sample with the selection criteria similar to those used by DESI (see Section~\ref{sec:ELG}). We present our results in Fig.~\ref{fig:ELG}. 
    
    As in previous cases, we can reproduce the clustering measurements within statistical uncertainties. This opens up the remarkable possibility of using SHAMe-SF to analyse the clustering of observational samples. Our model would be capable of employing most of the scale range in the data, providing constraints on model parameters that could inform the galaxy-subhalo connection and constrain star formation physics. Additionally, it could place constraints on cosmological parameters in the highly-nonlinear regime.
    
   As a future challenge, we aim to test whether the model is flexible enough to replicate the clustering of subsamples selected directly based on the intensity of emission lines.

\subsection{About model freedom}
    \label{sec:Nparam}
    
SHAMe-SF has 8 free parameters that control the relationship between halo and subhalo properties and the expected SFR galaxies. Next, we discuss how the number of model parameters impacts the clustering predictions. More concretely, we investigate the performance of SHAMe-SF when one parameter is held fixed.

Using the SHAMe-SF emulator, we repeat our MCMCs analysis fixing each parameter to 4 values within the range of our emulator. We do this for $z = 0$ and $z = 1$ and the two higher number densities. To quantify the performance of each case, we compare the respective $\chi^2$ value with that of the best-fitting model when all parameters are varied simultaneously.

We found that SHAMe-SF performs similarly when fixing the parameters that control the SFR-$\vpeak$ relation ($\beta$, $\gamma$, $\sigma$), or those that regulate the quenching of satellite galaxies ($\alpha_0$ or $\alpha_{\rm exp}$). This held regardless of the specific values at which each parameter was fixed. We can understand this in terms of internal degeneracies. For instance, a large scatter in the SFR-$\vpeak$ relation would be equivalent to weak slopes. Additionally, we expect that the details of satellite quenching are not important, given that its dependence on subhalo properties was not detected in our RF analysis (c.f. Section~\ref{sec:RF_results}).

On the contrary, we found that freedom on the peak of the SFR-$\vpeak$ relation was crucial to obtain accurate predictions. This is consistent with our RF analysis (and our model implementation), which determined that a SFR selection can be regarded as a $\vpeak$ selection at first order. Similarly, we find that the halo mass where the quenching starts acting $M_{\rm crit}$, is important in delivering good fits, specially at small scales. This is because quenching processes mostly affect the abundance and distribution of satellites.

Finally, the semi-sorting parameter has a strong effect on the clustering when fixed to random values. Small variations on this parameter can even double the value of the $\chi^2$. The effect of $f_k$ is partly compensated by other parameters, mainly reducing the distribution width to minimise its importance, especially for higher number densities. Lower values of $\sigma$ allow for more freedom in $f_k$ since it reduces its effect on the final distribution.  

It is important to take into account that we only analyse two number densities, redshifts and the specific case of TNG300: even if we can obtain a good clustering with fewer parameters, this does not mean that the model will perform equally in every situation, or when the SFR physics deviates strongly from that in TNG. Thus, we recommend using the full model (with 8 parameters) as a default and evaluate possible restrictions based on the specific dataset to analyse.

\section{Other statistics}
\label{sec:otherstats}
	To further validate the physical soundness of our best-fitting models, we will explore whether they correctly predict other statistics not directly included in the fitting procedure. Namely, we investigate the halo occupation number and magnitude of the assembly bias. This will test whether SHAMe-SF is selecting subhalos at the correct host halo mass and large-scale environments.
    
\subsection{HOD}
    \label{sec:HOD}

\begin{figure*}
		\centering
		\includegraphics[width=0.32\textwidth]{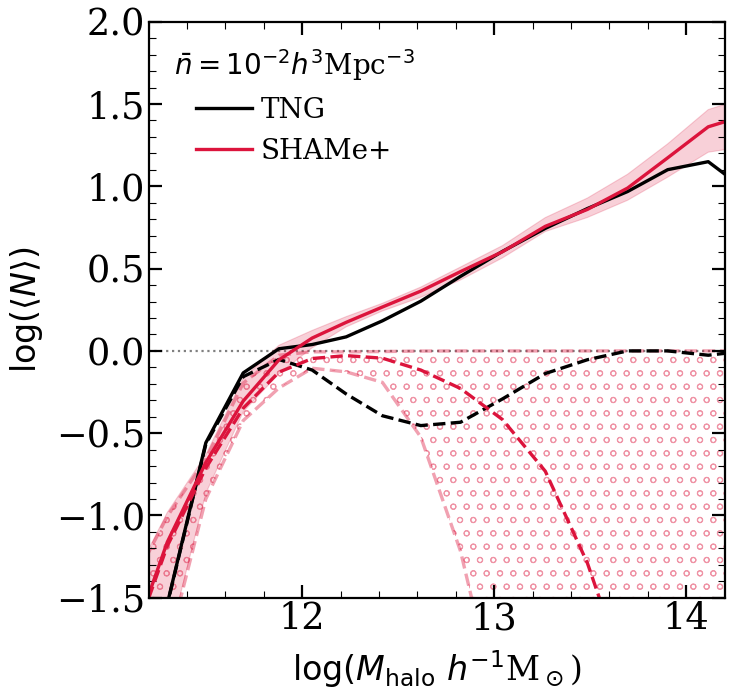}
        \includegraphics[width=0.32\textwidth]{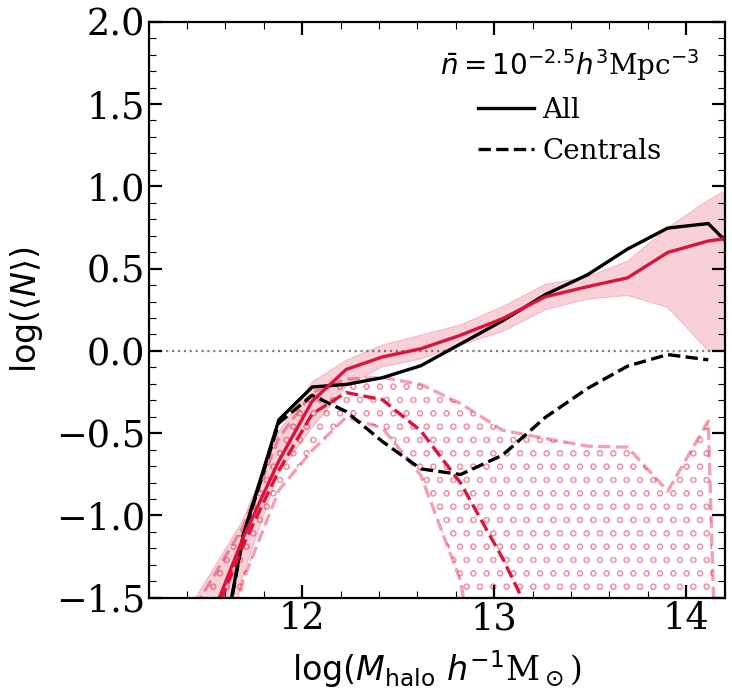}
        \includegraphics[width=0.32\textwidth]{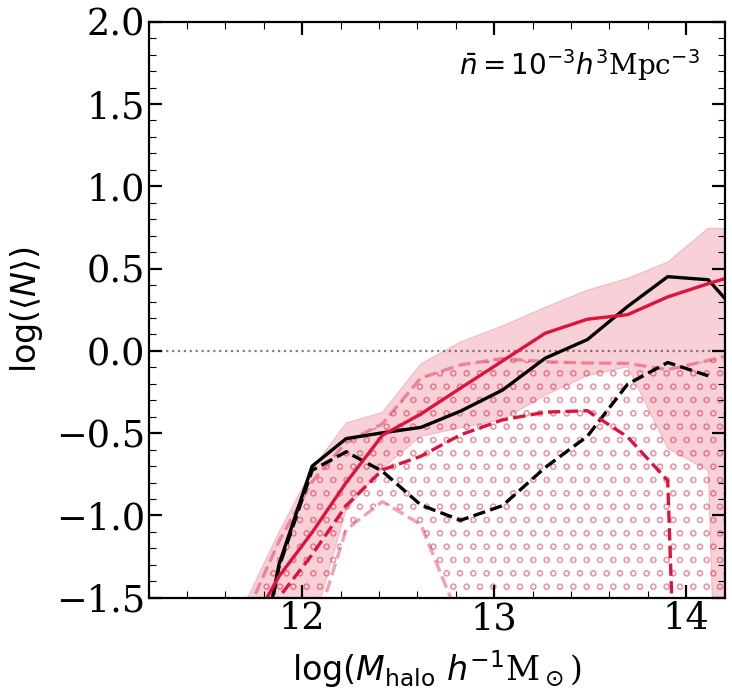}
		\caption{Halo occupation distribution (HOD) for galaxies in TNG300 (black) and the SHAMe-SF mocks (red) for $z = 0$ and number densities $n = 10^{-2}$, $10^{-2.5}$ and $10^{-3}\ihMpcC$. SHAMe-SF parameters were fitted to reproduce the clustering predictions. The shaded region represents the $1\sigma$ confidence interval (solid for centrals and galaxies, circle-texture for centrals).}
		\label{fig:HOD}
			
	\end{figure*}

    In Fig.~\ref{fig:HOD}, we present the average number of galaxies as a function of host halo mass. We display the total and central galaxies as solid and dashed lines, respectively. We show results for samples at $z=1$ as measured in TNG300 and in the corresponding best-fitting SHAMe-SF model. To illustrate the impact of parameter uncertainties in the best-fitting SHAMe-SF model, shaded regions indicate the range that encompasses the results for 200 parameter sets randomly selected within a $1\sigma$ interval in our MCMC chains.
    
    In all cases, SHAMe-SF predicts a distribution of host halo masses, for both central and satellites, that is very similar to that in the TNG300. In particular, satellite fractions, shown in Table~\ref{tab:satfractions}, are also remarkably similar. In the Appendix~\ref{app:clusteringmodel}, we show that SHAMe-SF yields a similar level of agreement at $z=0$ and when fitting the SAM catalogue.
    
    Perhaps an exception is the abundance of central galaxies in halos above $\sim 10^{13}\,\hMsun$, where SHAMe-SF predicts almost no star-forming galaxies. This is a consequence of SFR decreasing monotonically at large $\vpeak$, in contrast, TNG300 shows a ``rejuvenation'' of massive galaxies (c.f. Fig.~\ref{fig:ELG}). To investigate this further, we tested an extension of SHAMe-SF with a 2nd turning point allowing high $\vpeak$ subhalos to host star-forming galaxies even if intermediate $\vpeak$ subhalos are quenched. Including this extra freedom allowed a higher number of centrals at high halo masses but this regime remained largely unconstrained due to its minor impact on clustering statistics. However, it is worth considering this extension for specific statistics that would be sensitive to that regime.

	\begin{table}
	\begin{center}
    \begin{tabular}{ccc}
\textbf{Number density} & \textbf{TNG300} & \textbf{SHAMe-SF} \\ 
$[\ihMpcC]$ & $f_{\rm sat}$ & $f_{\rm sat}$ \\
\hline
$10^{-2}$   & $0.33$  &  $0.35\pm0.02$ \\
$10^{-2.5}$ & $0.32$  &  $0.40\pm0.02$ \\
$10^{-3}$   & $0.33$  &  $0.38^{+0.04}_{-0.03}$
\end{tabular}
    \caption{Satellite fractions for the three highest number density samples in TNG300 at $z=1$ and the respective SHAMe-SF predictions.}
    \label{tab:satfractions}
    	\end{center}
    \end{table}
	
	\subsection{Assembly bias}
 \label{sec:clusbias}

The final statistic we explore is the magnitude of the so-called ``assembly bias''.  Assembly bias quantifies the dependency of large-scale clustering on halo properties other than mass \citep{Sheth:2004, Gao:2005, Gao:2007, Wechsler:2006, Faltenbacher:2010, Angulo:2009, Mao:2018}. The galaxy-halo connection implies that this effect can be generalised to galaxies, defining a "galaxy assembly bias".
   
    To measure the degree of assembly bias in our samples, we use the technique proposed by \cite{Croton:2007}. In this approach, the positions of halos (including all their satellites) are shuffled among halos within $0.1$ bins in halo mass. Then, assembly bias is estimated by comparing the shuffled and original correlation functions:
    \begin{equation}
    b^2 = \xi(r)/\xi_{\rm shuffled}(r).
    \end{equation}
   
    For SFR-selected samples, assembly bias decreases for lower number densities at larger $r$, even becoming "negative" \citep[lower than the shuffled sample,][]{C19}. However, we note that the amount of galaxy assembly bias differs for different models and schemes \citep{Croton:2007, ChavesMontero:2016}, owing to the assumed impact of secondary halo properties on a galaxy SFR. 
    
\begin{figure*}
		\centering
		\includegraphics[width=0.47\textwidth]{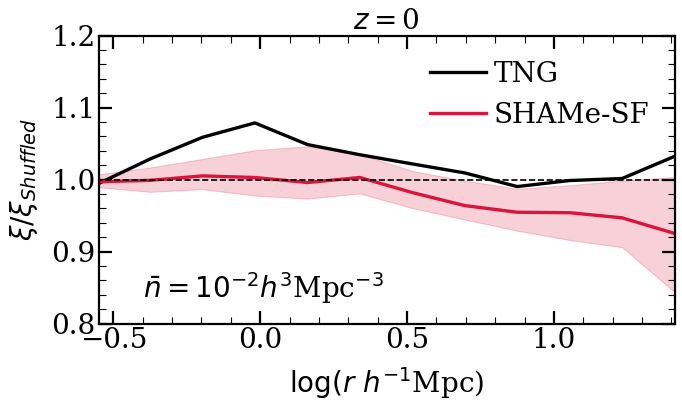}
        \includegraphics[width=0.47\textwidth]{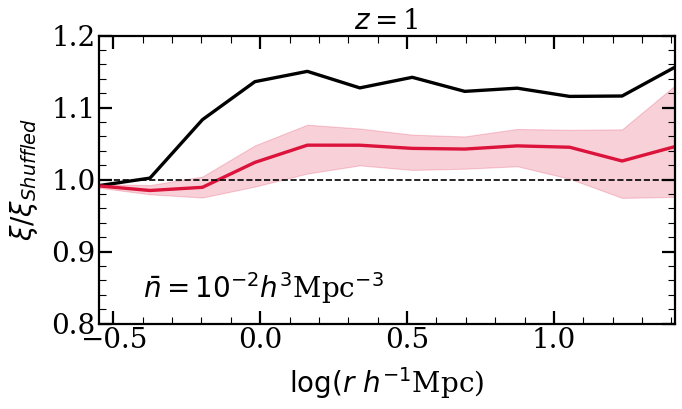}

        \includegraphics[width=0.47\textwidth]{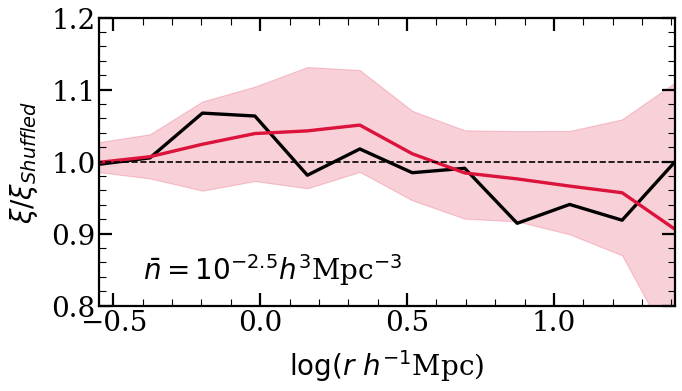}
        \includegraphics[width=0.47\textwidth]{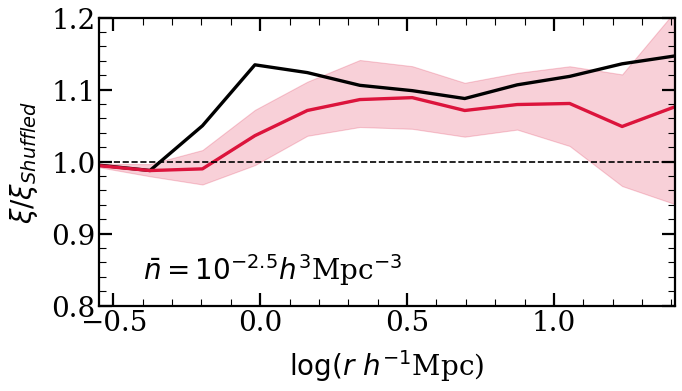}
		\caption{Galaxy assembly bias for galaxies in TNG300 (black) and the SHAMe-SF mocks for $z = 0$ (\textbf{left}) and $z = 1$ (\textbf{right}), and number densities $n = 10^{-2}$ (\textbf{top}) and $10^{-2.5}$ $\ihMpcC$ (\textbf{bottom}). SHAMe-SF parameters were fitted to reproduce the clustering predictions (red). The shaded region represents the $1\sigma$ confidence interval.}
		\label{fig:assemblybias}
	\end{figure*}

    Results for the two highest number densities samples at $z = 0$ and $z = 1$ from the TNG300 and the corresponding SHAMe-SF are shown in  Fig.~\ref{fig:assemblybias}. We show the average over 10 random shuffles to reduce statistical noise. As in the previous sections, we estimate the role of uncertainty in the best fits by considering 200 parameter sets within a $1\sigma$ region in our chains. 

For the sparser sample, we obtain a behaviour statistically consistent with TNG300 at both $z=0$ and $1$. Specifically, at $z=1$ the amount of assembly bias changes the clustering by approximately $10\%$ in both TNG300 and SHAMe-SF. Even if we did not include any parameter to explicitly model assembly bias (see \citealt{Contreras:2021shame}). However, for the denser sample, we slightly overestimate the amount of assembly bias. It is interesting to note that both samples at $z=0$ display a scale-dependent assembly bias, which tends to reduce the clustering amplitude, as also previously reported by \cite{C19} \citep[see also][]{Jimenez:2021}. By applying SHAMe-SF to larger simulations, we will be able to explore this with better statistical precision.
	
	\section{Summary}
	\label{sec:conclusions}
In this study, we introduce SHAMe-SF, a physically motivated model designed to populate gravity-only simulations, capable of predicting the clustering of star-formation-rate (SFR)-selected samples across various number densities and redshifts. 

	Our initial step involved utilizing a Random Forest algorithm to explore the connection between dark matter structures and SFR in the TNG300 simulations. Although star formation is a highly complex and potentially stochastic process, we showed that the clustering of SFR-selected samples can be predicted using only dark matter properties.
      
    After identifying the most relevant subhalo properties, we constructed SHAMe-SF. Our model involves the following ingredients:

    \begin{itemize}
\item Modeling the SFR-$\vpeak$ relationship as a broken power law with three free parameters (Eq.~\ref{eq:SFR}).
\item Introducing two additional parameters to represent deviations from the mean relation, controlling the magnitude of the scatter and its correlation with $\vpeak/\vvir$.
\item Modulating the SFR of satellite galaxies with three additional parameters, which reduce the SFR associated with a subhalo based on the time elapsed since reaching its peak mass and the host halo's mass.
    \end{itemize}
    
 Subsequently, we trained a neural network that emulates the SHAMe-SF clustering predictions in less than 1 second of CPU time. With this emulator, we showed that SHAMe-SF reproduces the clustering of various mock galaxy samples. Remarkably, this was achieved using a gravity-only simulation with 64 times lower resolution than TNG300, and over a broad range of scales $r \in [0.6, 30]\hMpc$. Specifically, the key milestones can be summarised as follows:
 
    \begin{itemize}
        \item SHAMe-SF accurately describes the projected correlation function, monopole and quadrupole of the redshift-space correlation function of TNG300 samples,
three number densities ($\bar{n}[\ihMpcC] = \{10^{-2},10^{-2.5}, 10^{-3}, 10^{-3.5}\}$) and two redshifts ($z = 0$ and $z = 1$) (Fig.~\ref{fig:clusteringz1}).
        \item We demonstrated the flexibility of SHAMe-SF by fitting the clustering of samples predicted by \texttt{LGalaxies}, a semi-analytical model which adopts different galaxy and star formation prescriptions than TNG300 (Fig.~\ref{fig:SAMz1}).
        \item Additionally, SHAMe-SF also was able to describe the clustering of mock ELG galaxies at $z=1$, which were selected using photometric criteria analogous to those employed y the ongoing DESI survey (Fig.~\ref{fig:ELG}).
        \item We validated the physical realism of our best-fitting SHAMe-SF models by comparing the mean occupation number (Fig.~\ref{fig:HOD}) and the magnitude of the so-called assembly bias. (Fig.~\ref{fig:assemblybias}). SHAMe-SF retrieves trends in qualitative agreement with those in the mock samples.  
    \end{itemize}

  Recently, various models have been developed specifically ELGs using approaches like the HODs and SHAM extensions (e.g., \citealt{Lin:2023}, [OII] in eBOSS; \citealt{Favole:2017}, [OII] in SDSS; \citealt{Yu:2023,Prada:2023}, ELGs in DESI). Although our work primarily focused on SFR selections, the flexibility of SHAMe-SF, coupled with the strong correlation between SFR and emission lines, suggests its applicability to ELGs. We will explore this in the future. This prospect becomes particularly interesting when combined with suites of $N$-body simulations or re-scaled simulations \citep{Zennaro:2019, C20, Angulo:2021}, which would offer opportunities to constrain cosmological parameters with future galaxy surveys from highly nonlinear scales.
	
\section*{Acknowledgements}
We would like to thank Jon\'{a}s Chaves-Montero, Idit Zehavi and Violeta Gonz\'{a}lez-P\'{e}rez for useful comments on the manuscript and the project. The authors acknowledge support by the project PID2021-128338NB-I00 from the Spanish Ministry of Science. SOM is funded by the Spanish Ministry of Science and Innovation under grant number PRE2020-095788. SC acknowledges the support of the ``Juan de la Cierva Incorporac\'ion'' fellowship (IJC2020-045705-I).
The authors also acknowledge the computer resources at MareNostrum and the technical support provided by Barcelona Supercomputing Center (RES-AECT-2019-2-0012 \& RES-AECT-2020-3-0014). The first Random Forest calculations followed the example provided by Saurabh Kumar in \citep{Kumar:2021} \footnote{\url{https://saurabhkumar3400.com/assemblybias.html}}. We further thank the developers of the open-source tools used in this work:  \texttt{matplotlib} \citep{Hunter:2007}, \texttt{numpy} \citep{Walt:2011}, \texttt{scipy} \citep{Jones:2001}, \texttt{scikit-learn} \citep{Pedregosa:2011}, \texttt{pandas} \citep{mckinney-proc-scipy-2010} and \texttt{emcee} \citep{ForemanMackey:2013}.
	\section*{Data Availability}
The IllustrisTNG simulations, including TNG300, are publicly available and accessible at \url{www.tng-project.org/data} \citep{Nelson:2019a}. The data underlying this article will be shared on reasonable request to the corresponding author.

	
	
	\bibliographystyle{mnras}
	\bibliography{example} 

	
	
	
	\appendix

    \section{Computing clustering statistics}
    \label{sec:AppClustering}
	
The main galaxy clustering statistics computed in this work are the projected correlation function and two multipoles (monopole and quadrupole) of the correlation function. When we analyse the clustering predicted by the RF models, we use the cross-correlation function between the test side of the box and the full volume. When analysing the predictions of the model, we use the auto-correlation function. 

To compute $w_p$, we use \textsc{corrfunc} \citep{Corrfunc1, Corrfunc2} integrating the auto-correlation function over the line of sight following:
\begin{equation}
    w_{\rm p} = 2 \times \int_{0}^{\pi_{\rm max}} \xi(\rm p, \pi)d\pi,
\end{equation}
where $\xi(\rm p, \pi)$ is the two-point correlation function. We choose $\pi_{\rm max}=30\hMpc$ as the maximum depth used on the pair counting due to the small size of the TNG300 simulation box.

To obtain the multipoles, we first compute the correlation function in terms of $s$ and $\mu$ bins defined as: $s^2 = r^2_{\rm p} + r^2_{\pi}$ and $\mu = \cos\left(\hat{s,\rm los} \right)$. Then, we define the multipoles as:
\begin{equation}
    \xi_{\ell} = \frac{2 \ell+1}{2} \int^{1}_{-1} \xi(s,\mu)P_\ell(\mu)d\mu,
\end{equation}
\noindent where $P_{\ell}$ is the $\ell$ -th order Legendre polynomial. We also use \textsc{corrfunc} to compute these statistics.
	
    \section{RF models $\chi^2$}
    \label{app:chi2}
\begin{figure}
		\centering
		\includegraphics[width=0.45\textwidth]{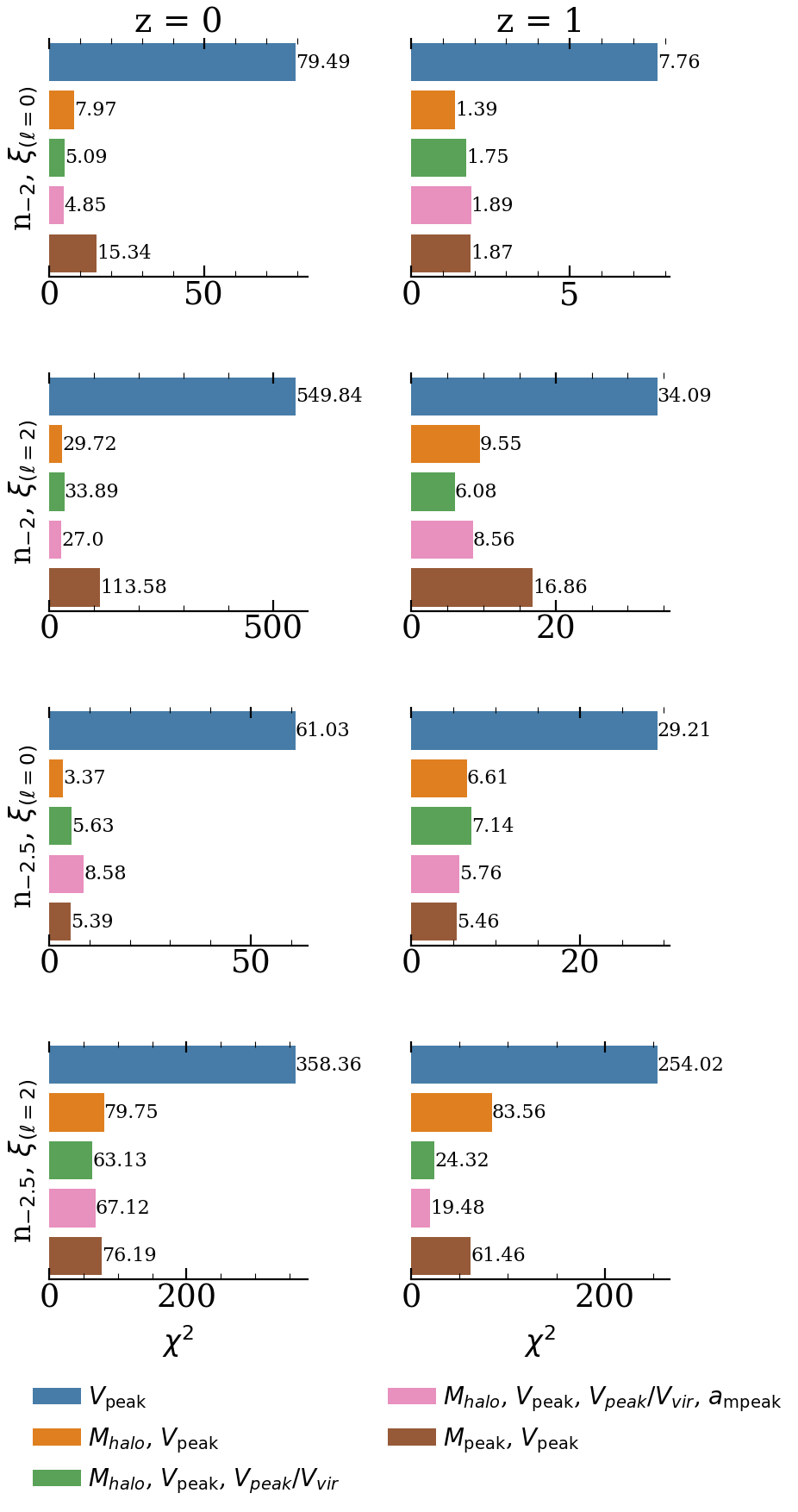}
		\caption{$\chi^2$ values of three different models computed using RF with different number of input properties (colour coding). We show the results for the monopole and quadrupole for number densities, $\bar{n}\,[\ihMpcC] = \{10^{-2},10^{-2.5}\}$ at two different redshits (columns, $z = 0$ and $z = 1$). More information about how the $\chi^2$ was computed can be found in Section~\ref{sec:RF_results}.}
    \label{fig:RFstat}
	\end{figure}

    Here, we show the $\chi^2$ of some of the models trained with RF. Details on the samples and the calculation were discussed in Section~\ref{sec:RFchi}. We compare the measurements of the cross-correlation between the validation sample (half of the box) and the complete sample for TNG300 and the RF models. This choice reduces the statistical noise while reducing the impact of the training set. However, to compute the contribution of the jackknife samples to the covariance matrix ($C_v$), we use the auto-correlation function of the full galaxy sample. This choice will not affect the model comparison since we use the same covariance matrix, but we expect the values of the $\chi^2$ to be higher due to the size of the errors. 

    In Fig.~\ref{fig:RFstat} we show the $\chi^2$ values of the monopole and the quadrupole for two number densities, two redshifts and five models trained using different properties. We show the models from Fig.~\ref{fig:RFresult}, the final property choice of SHAMe-SF ($\vpeak$, $\mhalo$, $\vpeak/\vvir$, $\tmpeak$) and the model trained with $\vpeak$ and $\mpeak$ (as an example of another model with two properties). As discussed in Section~\ref{sec:RFplots}, the most significant improvement appears when adding the second property. Adding the third (or fourth) property can enhance the prediction for some statistics (see the quadrupole for the highest number density at both redshifts), producing small noise-driven fluctuations in the $\chi^2$ of other statistics for the same property choice.

 	\section{TNG300-SHAMe-SF clustering at $z = 0$}

	\begin{table}
	\begin{center}
    \begin{tabular}{ccc}
\textbf{Number density} & \textbf{TNG300} & \textbf{SHAMe-SF} \\ 
$[\ihMpcC]$ & $f_{\rm sat}$ & $f_{\rm sat}$ \\
\hline
$10^{-2}$   & $0.34$  &  $0.47\pm0.07$ \\
$10^{-2.5}$ & $0.32$  &  $0.41\pm^{+0.13}_{-0.10}$ \\
$10^{-3}$   & $0.34$  &  $0.53^{+0.05}_{-0.13}$
\end{tabular}
    \caption{Satellite fractions for the two higher number densities for TNG300 and the prediction of the SHAMe-SF model for $z = 0$.}
    \label{tab:satfractionsz0}
    \end{center}
    \end{table}
    
\begin{figure*}
    		\centering
    		\includegraphics[width=0.33\textwidth, height = 0.26\textheight]{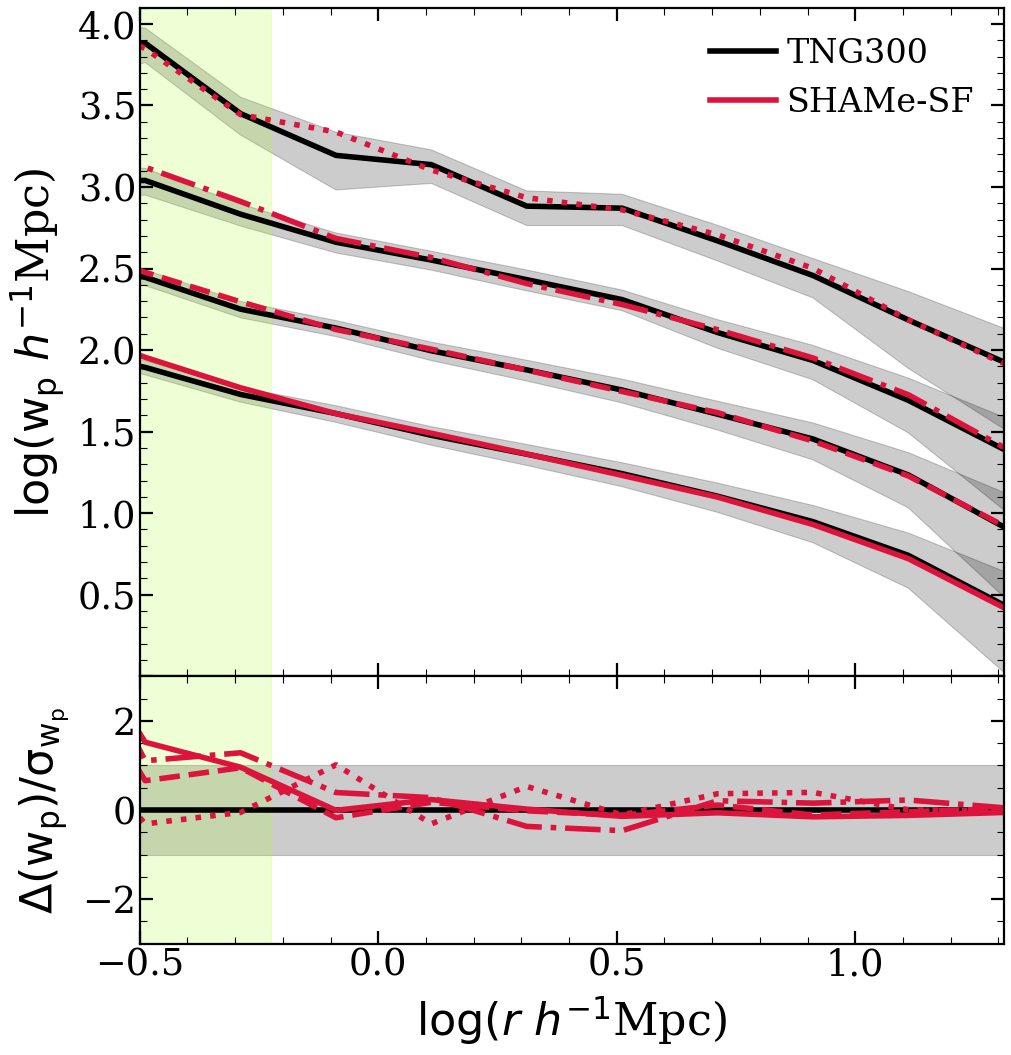}
            \includegraphics[width=0.33\textwidth, height = 0.263\textheight]{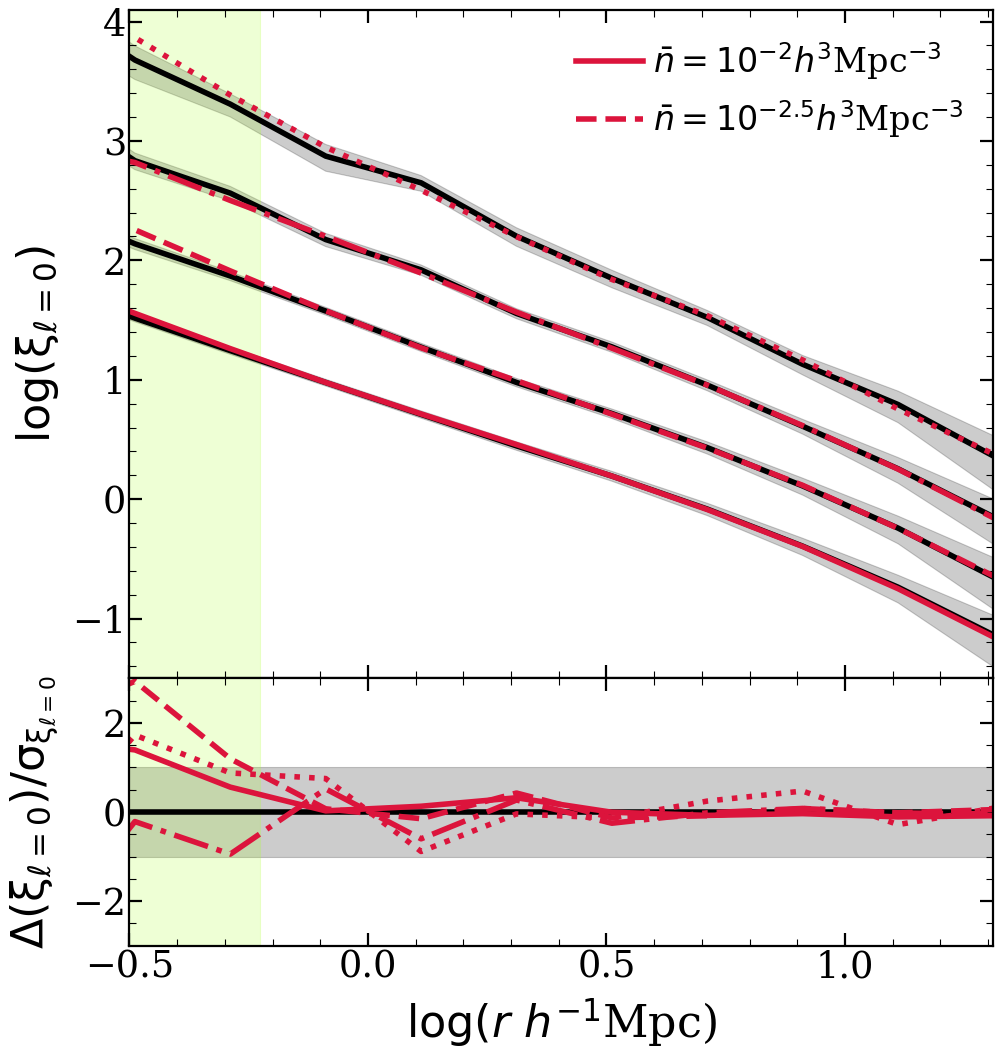}
            \includegraphics[width=0.33\textwidth, height = 0.26\textheight]{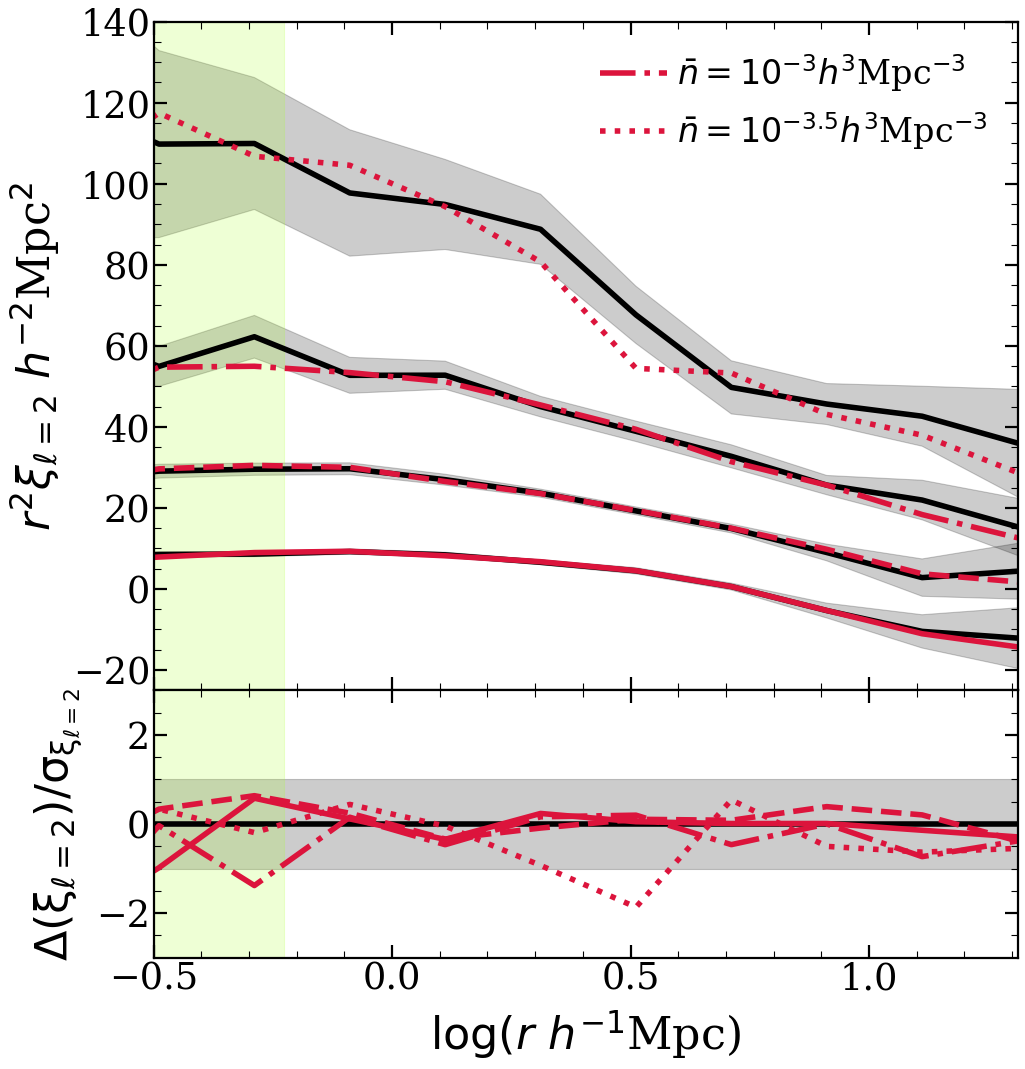}

            \includegraphics[width=0.32\textwidth]{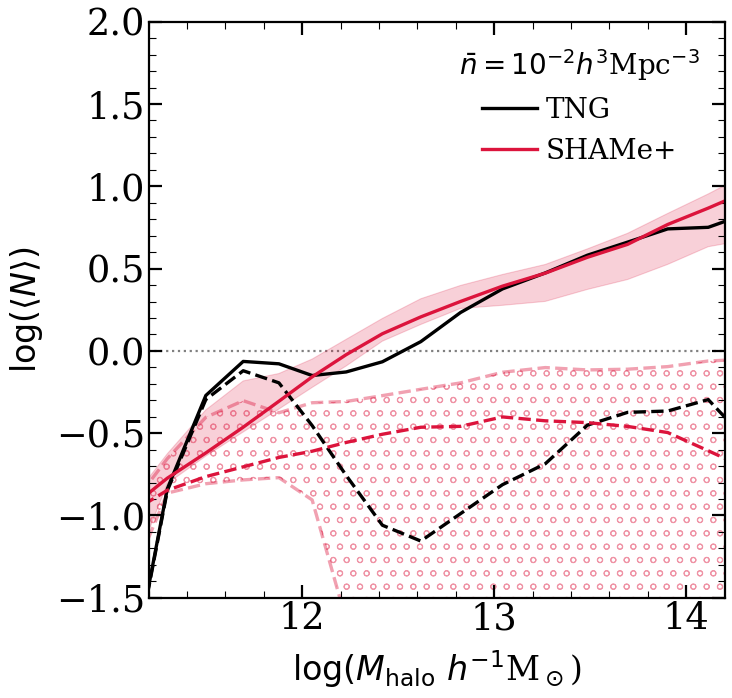}
        \includegraphics[width=0.32\textwidth]{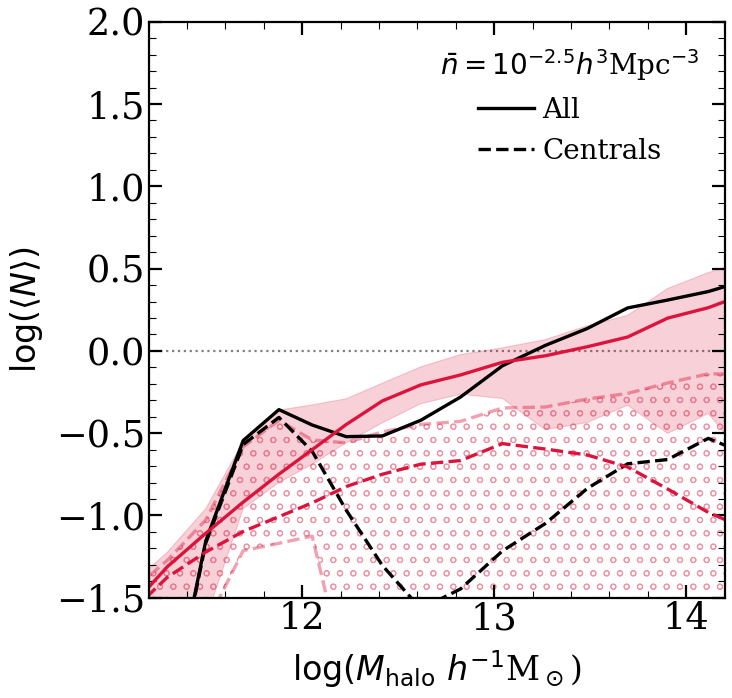}
        \includegraphics[width=0.32\textwidth]{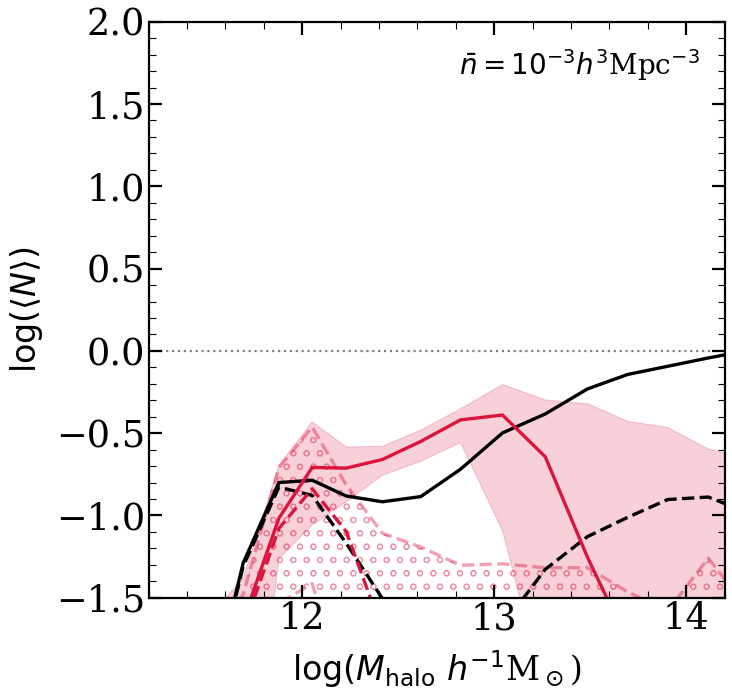}
    		\caption{Same as Figure~\ref{fig:clusteringz1}, but for galaxies (TNG300 and mock sample) at $z = 0$. }
    		\label{fig:clusteringz0}
	\end{figure*}

  Even if future surveys will target star-forming galaxies at higher redshifts ($z\sim 1$), we also test the clustering predictions of SHAMe-SF at $z = 0$. We show the results in Fig.~\ref{fig:clusteringz0} for the projected correlation function, monopole and quadrupole (upper panel) and the predictions of the HOD (lower panel). The performance of the clustering statistics used for the fit is similar to $z = 1$, discussed in Section~\ref{sec:clustering}. The main differences appear on the HOD, especially for high-mass central halos. As discussed in Section~\ref{sec:HOD}, the model cannot reproduce two turnarounds in the SFR function. This reflects on the satellite fractions (Table~\ref{tab:satfractionsz0}), which tend to be overestimated, mostly due to the lack of central subhalos.
  \label{app:clusteringmodel}

	\section{\texttt{L-Galaxies}}
    \label{sec:AppLgalaxies}
\begin{figure*}
		\centering
		\includegraphics[width=0.45\textwidth]{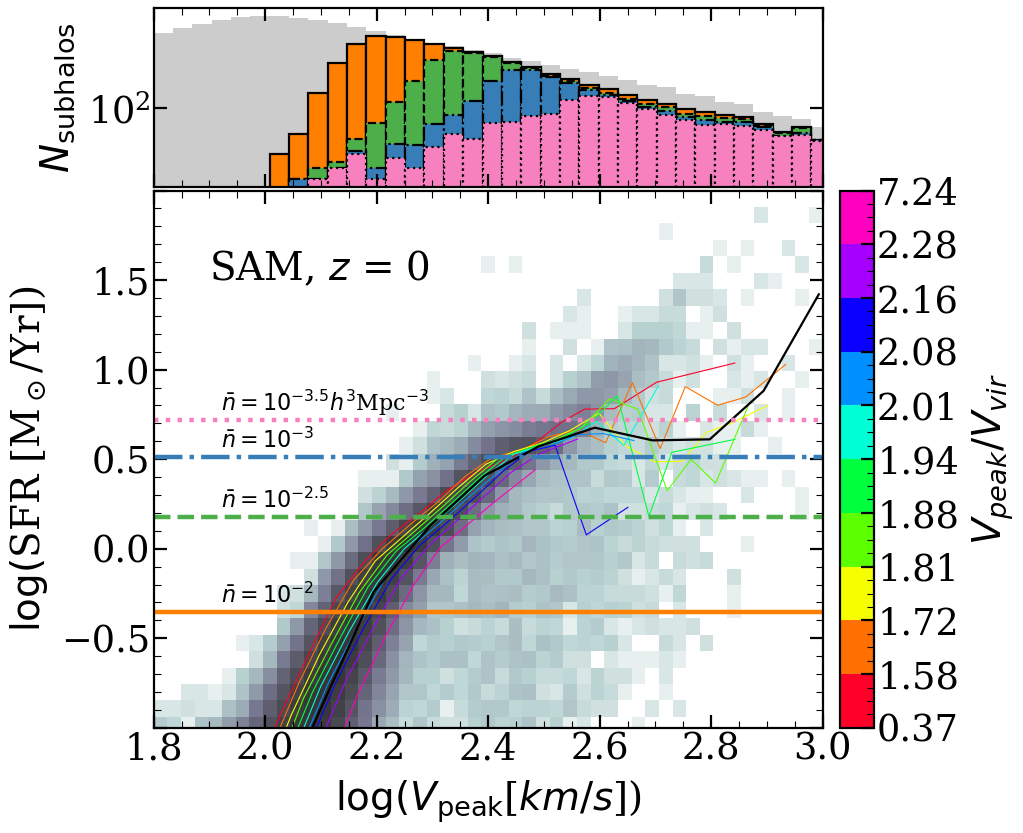}      \includegraphics[width=0.45\textwidth]{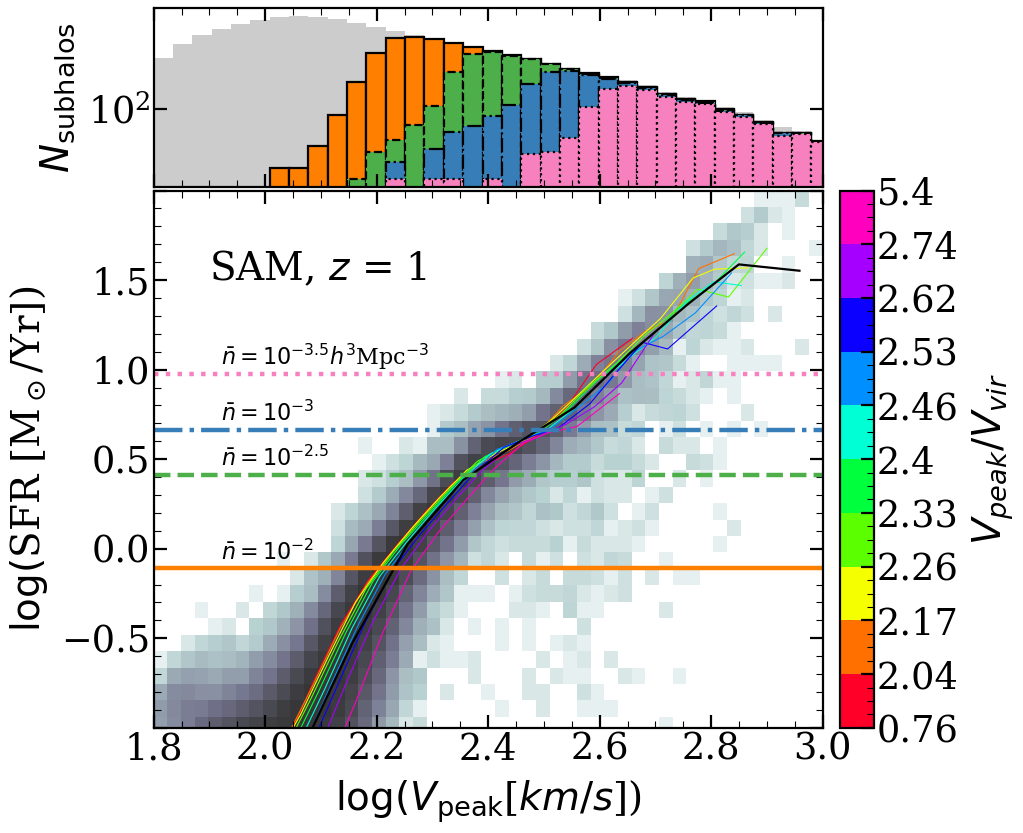}
		\caption{Same as the top and middle pannel of Figure~\ref{fig:vpeakSFR}, but for SAM galaxies at $z = 0$ (\textbf{left}) and 1 (\textbf{right}).}
        
		\label{fig:vpeakSFRLgal}
	\end{figure*}

    To test SHAMe-SF with an SFR prescription different from the TNG300, we use the SAM described in Section~\ref{sec:sam}. Before trying to fit the clustering, we prove that the main parametrization of the model (Section~\ref{sec:Model}) can also describe the relation between $\vpeak$ and SFR. We replicate the upper and centre panels of Fig.~\ref{fig:vpeakSFR} in Fig.~\ref{fig:vpeakSFRLgal} for $z = 0$ (left) and $z = 1$ (right). In the SAM model, the quenching mechanisms for high-mass halos are not as efficient as in TNG300, especially for $z = 1$. In this case, the most star-forming galaxies are hosted mostly by the subhalos with higher $\vpeak$ (as appreciable in the histograms for the higher number densities). As discussed in Section~\ref{sec:SAMcluster}, SHAMe-SF can describe this behaviour setting $\gamma$, the second slope in Eq.~\ref{eq:SFR}, to negative values. We note that we find the same behaviour with $\vpeak/vvir$, where less concentrated subhalos would host galaxies with higher star formation rates for a fixed value of $\vpeak$.

\begin{figure*}
    		\centering
    		\includegraphics[width=0.33\textwidth, height = 0.26\textheight]{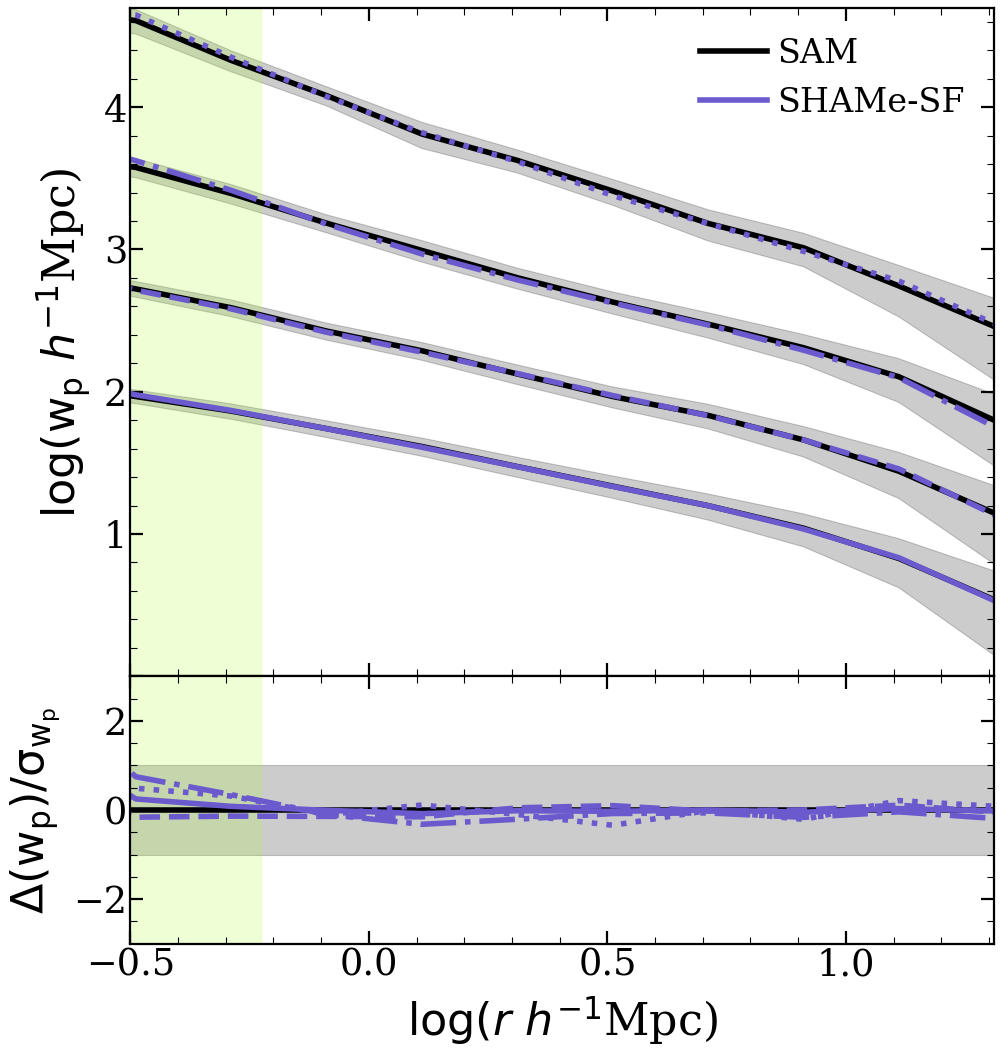}
            \includegraphics[width=0.33\textwidth, height = 0.263\textheight]{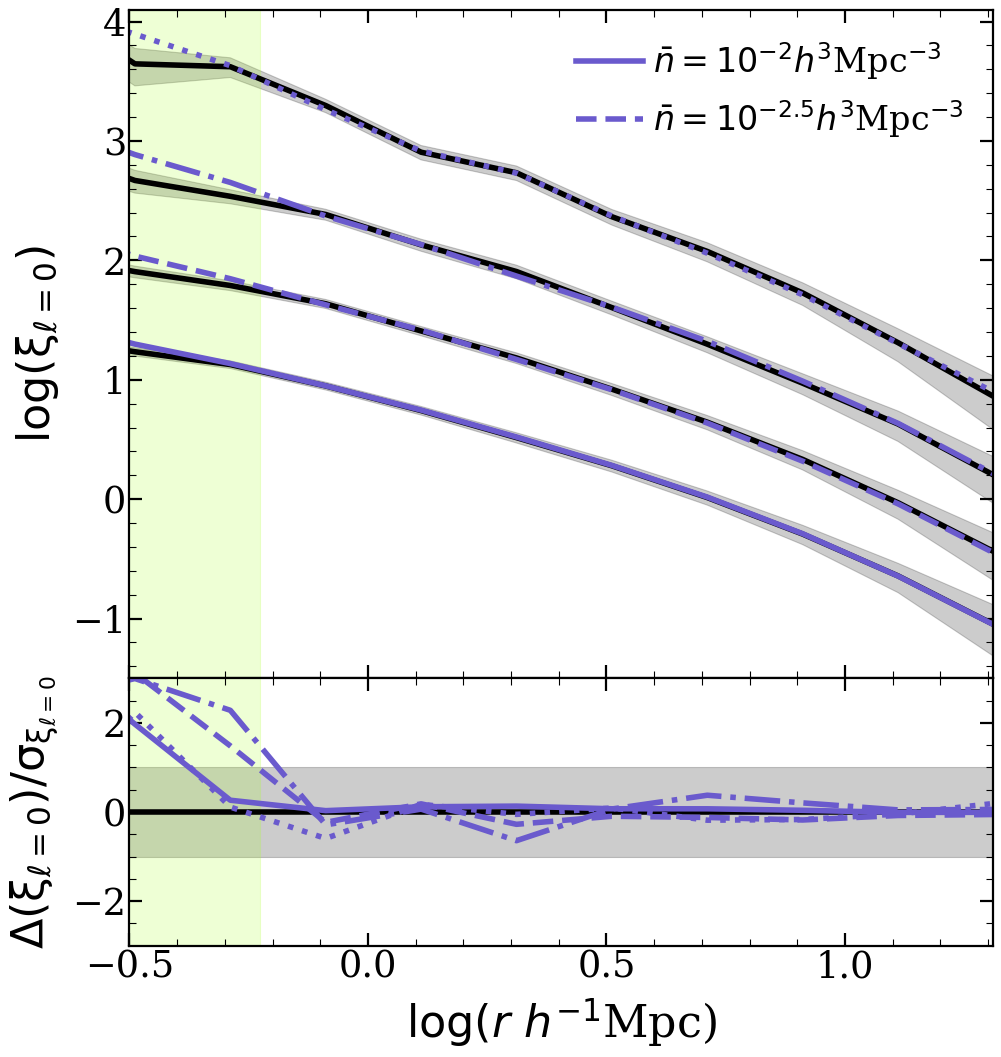}
            \includegraphics[width=0.33\textwidth, height = 0.26\textheight]{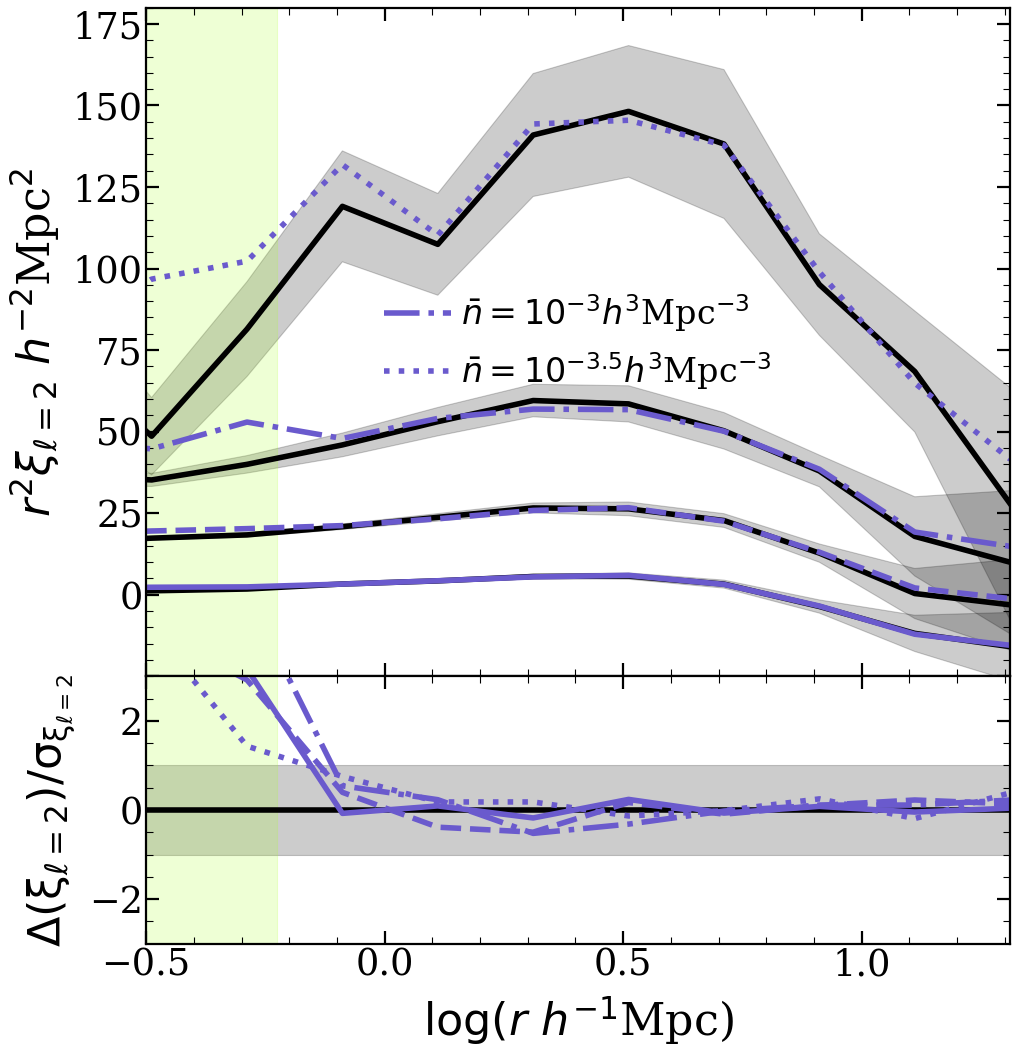}
    		\caption{Same as Figure~\ref{fig:clusteringz1}, but for SAM galaxies at $z = 0$. }
    		\label{fig:clusteringz0SAM}
	\end{figure*}

\begin{figure*}
		\centering
		\includegraphics[width=0.32\textwidth]{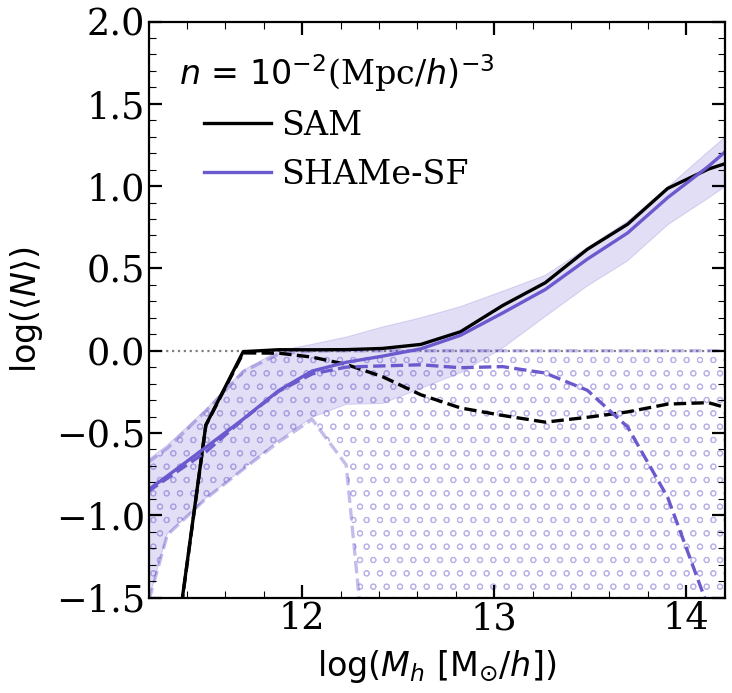}
        \includegraphics[width=0.32\textwidth]{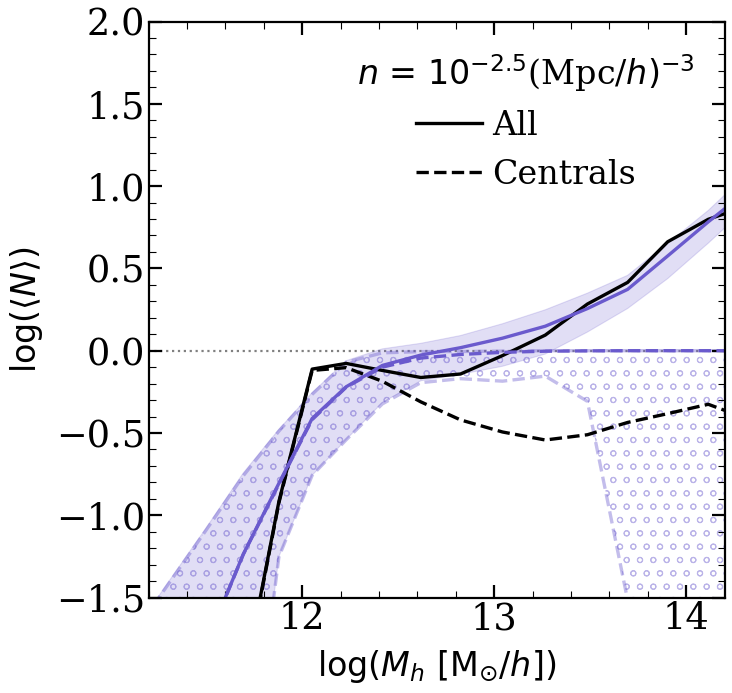}
        \includegraphics[width=0.32\textwidth]{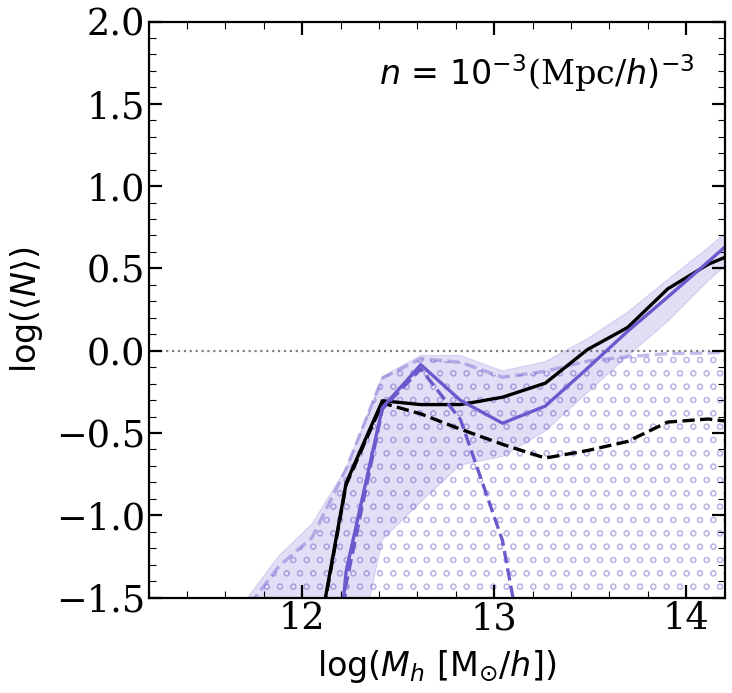}

        \includegraphics[width=0.32\textwidth]{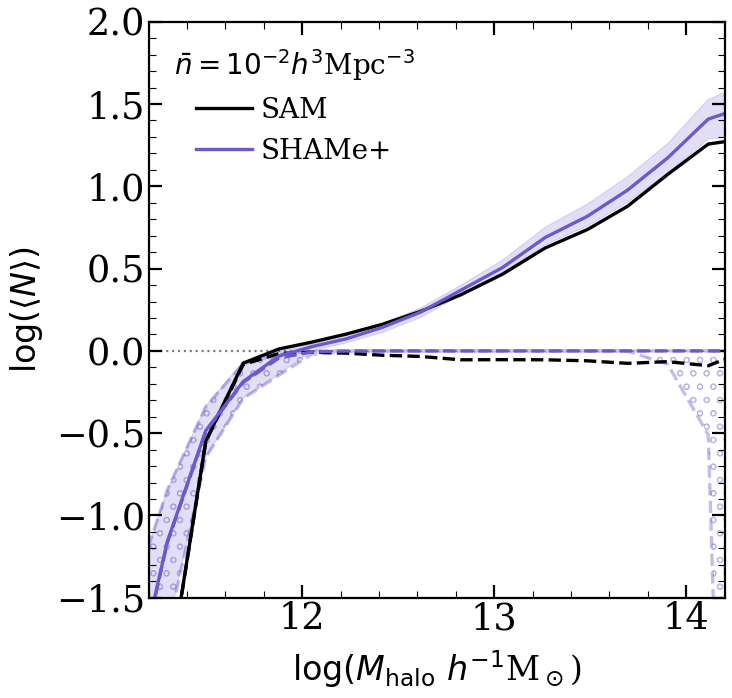}
        \includegraphics[width=0.32\textwidth]{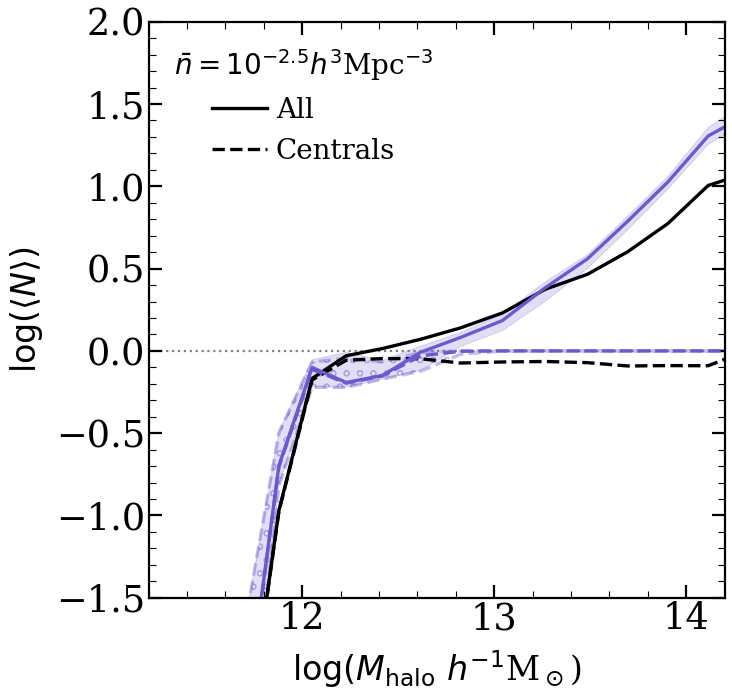}
        \includegraphics[width=0.32\textwidth]{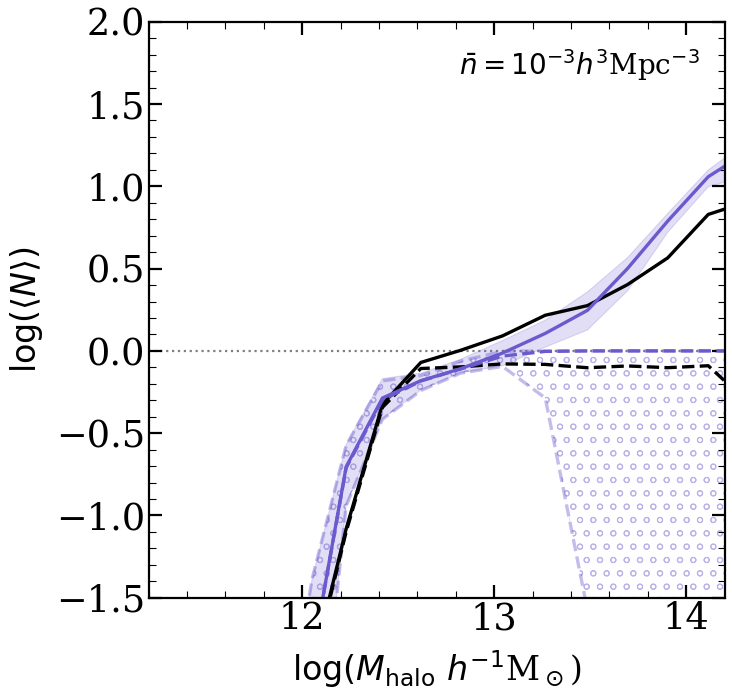}
		\caption{Halo occupation distribution (HOD) for galaxies in TNG300 (black) and the SHAMe-SF mocks (blue) for $z = 0$ (\textbf{top}) and $z = 1$ (\textbf{bottom}), and number densities $n = 10^{-2}$, $10^{-2.5}$ and $10^{-3}\ihMpcC$. SHAMe-SF parameters were fitted to reproduce the clustering predictions. The shaded region represents the $1\sigma$ confidence interval (circle-textured for centrals).}
		\label{fig:HODSAM}
	\end{figure*}

    To the clustering prediction shown for $z = 1$ in Fig.~\ref{fig:SAMz1}, we add the results for $z = 0$ in Fig.~\ref{fig:clusteringz0SAM}. We show the HODs for $z = 0$ (upper panel) and $z = 1$ (lower panel) in Fig.~\ref{fig:HODSAM}. 

    \

    \purple{ }
	
	\bsp	
	\label{lastpage}
\end{document}